\documentstyle[preprint,epsf,aps]{revtex}

\tightenlines

\draft

\begin{document}
\title{Spontaneous emission rate of an excited atom \\
placed near a nanofiber}
\author{V.V.Klimov}
\address{P.N.Lebedev Physical Institute, Russian Academy of Sciences, 53 Leninsky\\
Prospect, Moscow, 119991, Russia\\
E-mail: klimov@rim.phys.msu.su }
\author{M.Ducloy}
\address{Laboratoire de Physique des Lasers, UMR CNRS 7538\\
Institut Galilee, Universite Paris-Nord, \\
Avenue J-B. Clement F 93430 \\
Villetaneuse, France}
\maketitle

\begin{abstract}
The spontaneous decay rates of an excited atom placed near a dielectric
cylinder are investigated. A special attention is paid to the case when the
cylinder radius is small in comparison with radiation wavelength (nanofiber
or photonic wire). In this case, the analytical expressions of the
transition rates for different orientations of dipole are derived. It is
shown that the main contribution to decay rates is due to quasistatic
interaction of atom dipole momentum with nanofiber and the contributions of
guided modes are exponentially small. On the contrary, in the case when the
radius of fiber is only slightly less than radiation wavelength, the
influence of guided modes can be substantial. The results obtained are
compared with the case of dielectric nanospheroid and ideally conducting
wire.
\end{abstract}

\pacs{42.50.Pq, 33.50.-j, 32.80.-t, 12.20.Ds}

\section{Introduction}

\bigskip

At first sight, spontaneous emission is a purely atomic process. However as 
was first pointed out by Purcell [1], a resonant cavity can change the decay
rate substantially. At present it is known that not only resonant cavities
but also any material body can influence the decay rate of spontaneous
emission [2]. Moreover, the control of decay rate of spontaneous emission
 is widely used in practice of making the new efficient sources of
the light [3].

In spite of a qualitative understanding of the influence of material bodies on 
the spontaneous radiation of atom, only the influence of plane or spherical interface
 on decay rate has been elaborated in detail [4-10].

However, the influence of dielectric fiber or metallic wire on decay rate of
a single atom is interesting from a theoretical and practical point
of view. This is due to the fact that the charged wires are successfully
used to control atom motion [11,12]. The cylindrical geometry is also
important for investigation of fluorescence of substances in submicron
capillaries [13,14]. A very important application of the decay rate theory  in 
the presence of dielectric fiber is the photonic wire lasers [15,16]. Finally, cylindrical geometry appears naturally when considering carbon nanotubes [17,18].

The influence of ideally conducting cylindrical surface on decay rates is
well investigated both for atom inside a cylindrical cavity [19-21] and near
an ideally conducting cylinder [22].

The spontaneous emission of an atom in presence of a dielectric,
semiconductor or metallic cylinder is a more complicated process. First
investigation of the decay rate of an atom placed on the axis of dielectric
fiber was undertaken within classical approach in [23-24]. Recently that
problem attracted new interest [25-28]. In [29-30] spontaneous emission near
carbon nanotubes was considered. The general line of novel papers was a wide
using of numerical methods. Unfortunately such an approach cannot answer
many qualitative questions. In this paper, we re-investigate the influence
of a dielectric cylindrical surface on rates of dipole transitions, using
analytical and asymptotic approaches. For brevity we will consider only one
channel of decay. For example it may be 2P-1S channel or any other channel
of decay. To obtain clear analytical results we will investigate fiber with
radius small in comparison with  a wavelength. We will pay special attention to
 different mechanisms of the decay rate (radiative, waveguided, and
nonradiative).

The structure of the rest of the paper is as follows. In Section II , we
present general theory of decay rate in the presence of any body. We
shall also simplify general results in the case of nanobodies. In Sect. III, 
we consider the decay rates for an atom placed in a close vicinity to a nanofiber
 with any complex dielectric permittivity, when one can use quasistatic approximations.
 We obtain simple analytical expression for radiative and
nonradiative decay rates for any orientation of dipole momentum of atom. In
Sect. IV, we consider the full electrodynamic problem of dipole decay rate
near a nanofiber with any complex dielectric permittivity. We build the
analytical expression for decay rate through contour integral in complex
plane of longitudinal wave number. Then we transform general contour
integrals to separate contributions of guided and radiating modes. In
Section V, general expressions obtained in Section IV are applied to 
$z$ and $\varphi$, $\rho$ orientations of dipole momentum of an atom placed at the
surface of lossless fiber. In Section VI, we present graphical illustrations
and discuss the results obtained in previous Sections. Geometry of the
problem under investigation is shown in Fig.1. The dielectric permittivity
of a cylinder is $\varepsilon $, the dielectric permittivity of an outer space is equal to 1.

\section{Decay rate of an atom in the Presence of an Arbitrary Body}

In general the spontaneous emission of an atom placed in the vicinity of any
body is due to three different factors. First, the excitation energy can be
emitted in the form of photons moving to space infinity. It is natural to
name the corresponding part of decay rates as a radiative decay rate and to
designate it as $\gamma^{radiative}$. Second, the excitation energy can be
transformed to photons localized near or inside a body, that is, to guided
modes. We will denote the corresponding part of decay rates as $\gamma
^{guided}$. Finally, in the case of complex dielectric permittivity the
excitation energy can be transformed into thermal energy of a body. We will
denote a corresponding part of the decay rates as $\gamma^{nonradiative}$.

 Thus the total decay rate of an atom placed near any body can be
represented in the following form:

\begin{equation}  \label{eq1_1}
\gamma^{total} = \gamma^{radiative} + \gamma^{guided} + \gamma
^{nonradiative}
\end{equation}

In the case of excited molecules or quantum dots the total decay rate
includes also the internal nonradiative transitions. Such transitions have
no resonant nature and the influence of a body on these transitions seems to
be insignificant. In this paper we do not take such transition into account.

Let us now consider the radiative decay rate in more detail. Within
classical approach the excited atom can be described as a linear oscillator,
whose dipole momentum ${\rm {\bf d}}_{0}$ is proportional to matrix element
of dipole momentum operator. The oscillation frequency $\omega_{0}$
is equal to transition frequency. The radiation power of classical dipole in
free space is described by a well known expression \cite{Jackson}

\begin{equation}  \label{eq2_1}
\left( {{\frac{{dE}}{{dt}}}} \right)_{0}^{radiative} = {\frac{{c}}{{8\pi }}}%
\int {{\left| {\left( {{\rm {\bf E}}^{\left( {0} \right)}} \right)\times
\left( {{\rm {\bf H}}^{\left( {0} \right)}} \right)} \right|}_{r \to \infty
}^{2} r^{2}d\Omega} = {\frac{{ck^{4}}} {{3}}}{\left| {{\rm {\bf d}}_{0} }
\right|}^{2}
\end{equation}

\noindent where integration is over solid angle, ${\rm {\bf E}}^{\left({0}%
\right) },{\rm {\bf H}}^{\left( {0}\right) }$ are electric and magnetic
fields in free space, $k={\frac{{\omega }_{0}}{{c}}}$ stands for the wave
vectors in free space, and $c$ is the velocity of light in vacuum.

If one puts any body near excited atom, the radiation power will be changed
and will be described by the formula

\begin{equation}
\left( {{\frac{{dE}}{{dt}}}}\right) ^{radiative}={\frac{{c}}{{8\pi }}}\int {{%
\left| {\left( {{\rm {\bf E}}^{\left( {0}\right) }+{\rm {\bf E}}^{\left( {R}%
\right) }}\right) \times \left( {{\rm {\bf H}}^{\left( {0}\right) }+{\rm 
{\bf H}}^{\left( {R}\right) }}\right) }\right| }_{r\rightarrow \infty
}^{2}r^{2}d\Omega }  \label{eq3_1}
\end{equation}

\noindent where ${\rm {\bf E}}^{\left( {R} \right)}$ and ${\rm {\bf H}}%
^{\left( {R} \right)}$ are the electric and magnetic fields reflected by a
body.

If, additionally, the size of body (more precisely, the characteristic size
of the region with nonzero polarization) is small in comparison with
radiation wavelength and atom is placed near  a nanobody, the radiation of
a whole system will be of a dipole type. As a result, we will have, instead 
of (\ref{eq3_1}), the following expression

\begin{equation}
\left( {{\frac{{dE}}{{dt}}}}\right) ^{radiative}={\frac{{ck^{4}}}{{3}}}{%
\left| {{\rm {\bf d}}_{tot}}\right| }^{2}  \label{eq4_1}
\end{equation}

\noindent where ${\rm {\bf d}}_{tot}$ stands for total dipole momentum of
the whole system.

As the radiative decay rate is proportional to radiation power, the relative
radiative decay rate can be presented in the form:

\begin{equation}
{\frac{{\gamma ^{radiative}}}{{\gamma _{0}}}}={\frac{{{\left| {{\rm {\bf d}}%
_{tot}}\right| }^{2}}}{{{\left| {{\rm {\bf d}}_{0}}\right| }^{2}}}}
\label{eq5_1}
\end{equation}

\noindent where $\gamma _{0}$ is the decay rate of an excited atom in free
space.

Within quantum mechanical approach the radiative decay rate is described by
the Fermi golden rule \cite{Fermi}

\begin{equation}
\gamma ^{radiative}\propto {\left| {{\rm {\bf d}}_{0}{\rm {\bf E}}%
_{vac}\left( {{\rm {\bf r}}}^{\prime }\right) }\right| }^{2}\rho _{F}\left( {%
\omega }\right)   \label{eq6_1}
\end{equation}

\noindent where ${\bf d}_{0}$ is the  matrix element of dipole momentum
operator, ${\rm {\bf E}}_{vac}\left( {{\rm {\bf r}}}^{\prime }\right) $ is
the strength of electric field of emitted photon at the atom position, and 
$\rho _{F}\left( {\omega }\right) $ is the density of the final photon states.
Note that ${\rm {\bf E}}_{vac}\left( {{\rm {\bf r}}}^{\prime }\right) $ is
the solution of Maxwell's equations in free space. When the nanobody is present,
the radiative decay rate is described by the Fermi golden rule again \cite
{Fermi}

\begin{equation}
\gamma ^{radiative}\propto {\left| {{\rm {\bf d}}_{0}{\rm {\bf E}}\left( {%
{\rm {\bf r}}}^{\prime }\right) }\right| }^{2}\rho _{F}\left( {\omega }%
\right)   \label{eq7_1}
\end{equation}

\noindent but now ${\rm {\bf E}}\left( {{\rm {\bf r}}}^{\prime }\right) $ is
the solution of Maxwell's equations, which are modifies by the presence of
nanobody. The density of state $\rho _{F}\left( {\omega }\right) $ is still
independent of nanobody. Sometimes quantity ${\frac{{{\left| {%
{\rm {\bf d}}_{0}\cdot {\rm {\bf E}}\left( {{\rm {\bf r}}_{0}}\right) }%
\right| }^{2}}}{{{\left| {{\rm {\bf d}}_{0}}\right| }^{2}}}}\rho _{F}\left( {%
\omega }\right) $ is referred to as the radiative local density of states.
That quantity depends on the presence of nanobody.

As ${\rm {\bf E}}_{vac} \left( {{\rm {\bf r}}} \right)$ is nearly uniform on
the scale of nanobody, the nanobody acquires the dipole momentum $\delta 
{\rm {\bf d}} = \hat {\alpha} {\rm {\bf E}}_{vac} $, where $\hat {\alpha} $
is the nanobody polarizability tensor. As a result the influence of nanobody
on solution of Maxwell's equations can be described by the following
expression

\begin{equation}  \label{eq8_1}
{\rm {\bf E}} = {\rm {\bf E}}_{vac} + \delta {\rm {\bf E}},\quad \delta {\rm 
{\bf E}} = \hat {G}\delta {\rm {\bf d}}
\end{equation}

\noindent where $\delta {\rm {\bf E}}$ is the electric field of dipole with
momentum $\delta {\rm {\bf d}}$, and $\hat{G}$ is the tensor Green function of dipole source.
Using symmetry of $\hat{G}$ and $\hat{\alpha}$, it is
possible to show that

\begin{equation}
{\rm {\bf d}}_{0}{\rm {\bf E}}\left( {{\rm {\bf r}}}^{\prime }\right) ={\rm 
{\bf d}}_{total}{\rm {\bf E}}_{vac}\left( {{\rm {\bf r}}}^{\prime }\right) 
\label{eq9_1}
\end{equation}

\noindent where ${\rm {\bf d}}_{total}={\rm {\bf d}}_{0}+\hat{\alpha}\hat{G}%
{\rm {\bf d}}_{0}$ is the total dipole momentum of classical system where matrix
element ${\rm {\bf d}}_{0}$ is a dipole source. Now substituting (\ref{eq9_1}%
) into (\ref{eq7_1}) and averaging over polarizations of emitted photons we
again obtain expression (\ref{eq5_1}) for the relative radiative decay rate.

Thus, to find the radiative decay rate of an atom in the presence of
nanobody  it is sufficient to solve quasistatic problem and to find the total
dipole momentum of the whole system. Let us stress once more that (\ref
{eq5_1}) is valid within classical and quantum electrodynamics.

In the case of waveguided photons, the Fermi golden rule remains valid, but it is
impossible to obtain general expression for the case of nanobodies, because
the space structure of waveguided photons depends substantially on the nanobody geometry. 
The subsequent analysis (Section V) shows that the decay rate
into guided modes decreases exponentially with decreasing of nanobody size.

Let us now consider the nonradiative losses, that is,  losses due to presence
of imaginary part in dielectric permittivity (nonzero conductivity). In this
case one should use general expression for total decay rate of an atom near
any body \cite{Chance}:

\begin{equation}
{\frac{{\gamma }^{total}}{{\gamma _{0}}}}=1+{\frac{{3}}{{2}}}Im{\frac{{{\rm 
{\bf d}}_{0}{\rm {\bf E}}^{(R)}({\rm {\bf r}}^{\prime },\omega _{0})}}{{%
d_{0}^{2}k^{3}}}}  \label{eq10_1}
\end{equation}

Within this approach it is sufficient to find reflected field at the atom
position ${\rm {\bf E}}^{(R)}({\rm {\bf r}}^{\prime },\omega _{0})$. It is
very important that this expression is again valid within classical and 
quantum electrodynamics \cite{Wylie} - \cite{Welsch5}.

In general it is very difficult to find an analytical expression for
reflected field. However, in the case of nanobodies one can use perturbation
theory over wave vector $k$ (long wavelength perturbation theory)\cite
{Stevenson}. Within such theory the electric field can be presented by a series

\begin{equation}
{\frac{{{\rm {\bf d}}_{0}{\rm {\bf E}}^{(R)}({\rm {\bf r}}^{\prime },\omega
_{0})}}{{d_{0}^{2}}}}=a_{1}+b_{1}k+c_{1}k^{2}+id_{1}k^{3}+...  \label{eq11_1}
\end{equation}

\noindent where the coefficients $a_{{\rm 1}}$, $b_{{\rm 1}}$, $c_{{\rm 1}}$%
, and $d_{{\rm 1}}$ are determined by solving some quasistatic problems \cite
{Stevenson}. It is important to note that the first three terms are due to
near fields, while radiation fields appear only starting with the fourth
term, which is proportional to $k^{{\rm 3}}$. In the case of a medium with
losses, all of the coefficients $a_{{\rm 1}}$, $b_{{\rm 1}}$, $c_{{\rm 1}}$,
and $d_{{\rm 1}}$ are complex. Now substituting (\ref{eq11_1}) into (\ref
{eq10_1}) we obtain the series for the total decay rate:

\begin{equation}
{\frac{{\gamma }^{total}}{{\gamma _{0}}}}={\frac{{3}}{{2}}}%
\rm {Im} \left( \frac{a_{1}}{k^{3}}+\frac{b_{1}}{k^{2}}+\frac{c_{1}}{k}\right) +1+{%
\frac{{3}}{{2}}}%
\rm {Re} \left( d_{1}\right) +...  \label{eq12_1}
\end{equation}

For nonradiative part of decay rate from (\ref{eq1_1}) we have :

\begin{equation}  \label{eq13_1}
\gamma ^{nonradiative} = \gamma ^{total} - \gamma ^{radiative} - \gamma
^{guided}
\end{equation}

Due to the fact that the radiative and guided decay rates begin with the fourth term
of expansion, the nonradiative decay rate can be presented in the form :

\begin{equation}
{\frac{{\gamma }^{nonradiative}}{{\gamma _{0}}}}={\frac{{3}}{{2}}} {\rm Im}%
\left( \frac{a_{1}}{k^{3}}+...\right)  \label{eq14_1}
\end{equation}

Thus, to find the nonradiative decay rate it is sufficient to find
quasistatic reflected field. Let us stress once more that (\ref{eq14_1}) is
valid within classical and quantum approaches.

By and large, to find radiative and nonradiative decay rates in the case of
nanobodies it will suffice to solve quasistatic problem and to find the
reflected field and the total dipole momentum of the whole system. Of
course, the decay of an excited atom into waveguided modes can not be
described by quasistatic approximation.

\section{Quasistatic analysis of decay rate near nanofiber}

Let us consider now a decay rate of an excited atom placed near a nanofiber.
When the distance between the atom and fiber, and the radius of fiber are
substantially less than radiation wavelength, the fiber is polarized only
near the atom.

As a result, the radiation of atom + fiber system will be of the dipole-type
and can be described by (\ref{eq5_1}).

The total dipole momentum ${\bf d}_{{\rm t}{\rm o}{\rm t}}$ can be found
from the solution of quasistatic problem

\begin{equation}
\begin{array}{l}
rot{\rm {\bf E}}=0 \\ 
div{\rm {\bf D}}=4\pi \rho _{C}
\end{array}
\label{eq2}
\end{equation}

\noindent where the charge density $\rho_{C}$ at the point ${\bf r}$ is
derived from the dipole momentum of atom 

\begin{equation} \label{eq3}
\rho _{C}=-({\bf d}_{0}\nabla )\delta ^{(3)}({\bf r}-{\bf r}^{\prime
})e^{-i\omega t} ,  
\end{equation}

\noindent $\delta ^{(3)}({\bf r})$ is the three-dimensional Dirac delta function and ${%
\nabla }^{\prime }$ means gradient over radius vector of the atom, ${\bf r}%
{\rm {\bf ^{\prime }}}$. Hereafter we will omit the time dependence of the
fields. The continuity conditions for the tangential components of {\bf E}
and the normal components of {\bf D} at the surface of a cylinder with
dielectric constant $\varepsilon $ should be provided as well.

Introducing a potential $\tilde{\varphi}$ by

\begin{equation}  \label{eq4}
{\rm {\bf E}} = - \nabla \left( {{\rm {\bf d}}_{0} \nabla ^{\prime}} \right)%
\tilde {\varphi} \left( {{\rm {\bf r}},{\rm {\bf r}}^{\prime}} \right),
\end{equation}

\noindent we obtain, instead of (\ref{eq2}), the Poisson equation, 
\begin{equation}
{{\ 
\begin{array}{l}
{-\nabla ^{2}\tilde{\varphi}=4\pi \delta ^{(3)}({\bf r}-{\bf r}^{\prime
}),\quad \quad { \mbox{outside cylinder}}}  \\ 
{-\nabla ^{2}\tilde{\varphi}=0,\quad \quad { \mbox{inside cylinder}}}
\end{array}
}}  \label{eq5}
\end{equation}
It is convenient to represent the solution of problem ($\ref{eq5}$) in the
form

\begin{equation}
{{\ 
\begin{array}{l}
{\tilde{\varphi}=\tilde{\varphi}_{0}+\tilde{\varphi}_{2},\quad 
\mbox {outside cylinder}}\\ 
{\tilde{\varphi}=\tilde{\varphi}_{1},\quad \mbox{inside cylinder} }
\end{array}
}}  \label{eq6}
\end{equation}

\noindent where $\tilde{\varphi}_{{\rm 0}}$ is the free-space potential
given by

\begin{equation}
\tilde{\varphi}_{0}={\frac{{1}}{{{\left| {{\rm {\bf r}}-{\rm {\bf r}}%
^{\prime }}\right| }}}}  \label{eq7}
\end{equation}

This free space Green function can be expanded in cylindrical coordinates
system $\left( {\rho ,\varphi ,z}\right) $ (Fig.1) in the following series 
\cite{Smythe} :

\begin{equation}
{\frac{{1}}{{{\left| {{\rm {\bf r}}-{\rm {\bf r}}^{\prime }}\right| }}}}={%
\frac{{2}}{{\pi }}}{\sum\limits_{m=0}^{\infty }{\left( {2-\delta _{m,0}}%
\right) }}\cos m\left( {\varphi -{\varphi }^{\prime }}\right) {%
\int\limits_{0}^{\infty }{dh\cos h\left( {z-{z}^{\prime }}\right) }}%
K_{m}\left( {h{\rho }^{\prime }}\right) I_{m}\left( {h\rho }\right) \quad
\left( {\rho <{\rho }^{\prime }}\right)  \label{eq8}
\end{equation}

\noindent where $\delta _{m,0}$ is the Kronecker symbol, K and I are the modified
Bessel functions \cite{Abramowitz}.

Using this expansion we can found expressions for $\tilde{\varphi}_{{\rm 1}}$
and $\tilde{\varphi}_{{\rm 2}}$ by usual mode matching. For potential
outside fiber we have the following expression $\left( {\rho >{\rho }%
^{\prime }}\right) $

\begin{equation}  \label{eq9}
\tilde{\varphi}_{2}={\frac{{2}}{{\pi }}}{\sum\limits_{m=0}^{\infty }{\left( {%
2-\delta _{m}^{0}}\right) }}\cos m\left( {\varphi -{\varphi }^{\prime }}%
\right) {\int\limits_{0}^{\infty }{dh\cos h\left( {z-{z}^{\prime }}\right) }}%
K_{m}\left( {h{\rho }^{\prime }}\right) K_{m}\left( {h\rho }\right)
G_{m}\left( {ha}\right)
\end{equation}

\noindent where reflection coefficients $G_{{\rm m}}$ look as follows:

\begin{equation}  \label{eq10}
G_{m} \left( {s} \right) = {\frac{{\left( {\varepsilon - 1} \right){\frac{{%
dI_{m} \left( {s} \right)}}{{ds}}}I_{m} \left( {s} \right)}}{{{\frac{{dK_{m}
\left( {s} \right)}}{{ds}}}I_{m} \left( {s} \right) - \varepsilon {\frac{{%
dI_{m} \left( {s} \right)}}{{ds}}}K_{m} \left( {s} \right)}}}
\end{equation}

It is important to note that for negative values of dielectric constant, $%
Re(\varepsilon )<-1$ (metals), the denominator of (\ref{eq10}) is equal to
zero for some values of integration variable $h$. It means that under such
conditions, the guided modes do occur in the system.

To determine the total dipole momentum of the system one should find the
asymptotic value of (\ref{eq9}) at large distances, $\rho
^{2}+z^{2}\rightarrow \infty $. In the dielectric case $\left( 
\rm{Re}({\varepsilon )>1}\right) $ the main contribution to (\ref{eq9}) is due to
m=1 term and has the form

\begin{equation}
\tilde{\varphi}_{2}^{as}=-{\frac{{\varepsilon -1}}{{\varepsilon +1}}}{\frac{{%
a^{2}}}{{{\rho }^{\prime }}}}\cos \left( {\varphi -{\varphi }^{\prime }}%
\right) {\frac{{\rho }}{{\left( {\rho ^{2}+z^{2}}\right) ^{3/2}}}}
\label{eq11}
\end{equation}

Comparing the electric potential of reflected field

\begin{equation}
\tilde{\varphi}_{2}^{dip}=\left( {{\rm {\bf d}}_{0}{\nabla }^{\prime }}%
\right) \tilde{\varphi}_{2}^{as}  \label{eq12}
\end{equation}

\noindent with dipole potential

\begin{equation}
\tilde{\varphi}_{dip}={\frac{{{\rm {\bf d}}_{tot}{\rm {\bf r}}}}{{r^{3}}}}
\label{eq13}
\end{equation}

\noindent one can find the dipole momentum of fiber, ${\it \delta }{\bf d}$.
In the case of $\rho $-orientation of dipole momentum we have 
\begin{equation}
\delta d_{\rho }=d_{0,\rho}{\frac{{\varepsilon -1}}{{\varepsilon +1}}}{\frac{{%
a^{2}}}{{\rho }^{\prime 2}}}  \label{eq14}
\end{equation}

\noindent while in the case of $\varphi$-orientation the dipole momentum of
fiber will be

\begin{equation}
\delta d_{\varphi }=-d_{0,\varphi}{\frac{{\varepsilon -1}}{{\varepsilon +1}}}{\frac{{%
a^{2}}}{{\rho }^{\prime 2}}}  \label{eq15}
\end{equation}

In the case of z-orientation the dipole momentum induced in fiber is equal
to zero.

Now combining the dipole momenta of nanofiber and atom and substituting the
result in (\ref{eq5_1}) we obtain the radiative decay rates for an atom near
a nanofiber

\begin{equation}
\begin{array}{l}
\displaystyle\left( {{\frac{{\gamma }^{radiative}}{{\gamma _{0}}}}}\right)
_{\rho }={\left| {1+{\frac{{\varepsilon -1}}{{\varepsilon +1}}}{\frac{{a^{2}}%
}{{\rho }^{\prime 2}}}}\right| }^{2} \\ 
\displaystyle\left( {{\frac{{\gamma }^{radiative}}{{\gamma _{0}}}}}\right)
_{\varphi }={\left| {1-{\frac{{\varepsilon -1}}{{\varepsilon +1}}}{\frac{{%
a^{2}}}{{\rho }^{\prime 2}}}}\right| }^{2} \\ 
\displaystyle\left( {{\frac{{\gamma }^{radiative}}{{\gamma _{0}}}}}\right)
_{z}=1
\end{array}
\label{eq16}
\end{equation}

In the case of an atom placed at the fiber surface these results agree with
those obtained in the case of a prolate dielectric nanospheroid \cite{Klimov2,Klimov3}. 
Besides, these results agree with decay rates of an atom
placed on the axis of dielectric fiber \cite{Kats1,Kats2}. Indeed, for 
$\rho ^{\prime }=a$ we have from (\ref{eq16})

\begin{equation}
\left( {{\frac{{\gamma }^{radiative}}{{\gamma _{0}}}}}\right) _{\rho }={%
\left| {{\frac{{2\varepsilon }}{{\varepsilon +1}}}}\right| }^{2},\quad
\left( {{\frac{{\gamma }^{radiative}}{{\gamma _{0}}}}}\right) _{\varphi }={%
\left| {{\frac{{2}}{{\varepsilon +1}}}}\right| }^{2},\quad \left( {{\frac{{%
\gamma }^{radiative}}{{\gamma _{0}}}}}\right) _{z}=1  \label{eq17}
\end{equation}

\noindent while for the atom at the nanofiber axis one has \cite{Kats1,Kats2}.

\begin{equation}
\left( {{\frac{{\gamma }^{radiative}}{{\gamma _{0}}}}}\right) _{\rho
}=\left( {{\frac{{\gamma }^{radiative}}{{\gamma _{0}}}}}\right) _{\varphi }={%
\left| {{\frac{{2}}{{\varepsilon +1}}}}\right| }^{2},\quad \left( {{\frac{{%
\gamma }^{radiative}}{{\gamma _{0}}}}}\right) _{z}=1  \label{eq18}
\end{equation}

The only difference is for $\rho $- oriented dipole, where we have
enhancement by $\varepsilon ^{{\rm 2}}$. The physical interpretation is as
follows. From QED point of view, decay rate is due to interaction of the
dipole with electromagnetic modes, modified by a fiber. The tangential $%
(\varphi,z)$ components of mode electric field are continuous. So the decay
rates of dipoles with $(\varphi ,z)$ orientations at the surface and in the
body of the fiber should be the same. On the contrary, the $\rho$ - component of
mode electric field is discontinuous at the surface (the normal component of
\ $D=\varepsilon E$ is continuous). This explains the difference between decay
rates of $\rho$-oriented dipole inside and outside nanofiber.

We should also mention that our results do not agree with decay rates of an
atom inside photonic wire found in \cite{Chu1,Chu2}. We
believe that results \cite{Chu1,Chu2} are misleading because
of a very crude approximation of fiber by plane waveguide.

In the case of nanowires, that is, in the case of $Re(\varepsilon )<-1$, the
excitation of guided modes is possible and one should add additional (very
important !) terms to (\ref{eq16}). The detailed investigation of influence
of guided modes on decay rates near metallic nanowire will be presented
elsewhere \cite{Klimov4}.

Analogous additional terms occur in the case of dielectric fibers and they
are also due to the guided modes. In following Sections we will show that those
terms are exponentially small for nanofibers. As a result one can use
quasistatic formulae for the nanofiber safely.

The results obtained, (\ref{eq16}), are valid only for dielectric and
metallic nanocylinders. The situation is dramatically changed in the case of
an ideal conductor $\left( {{\left| {\varepsilon }\right| }\rightarrow \infty }%
\right) $, when reflection coefficients become equal to

\begin{equation}  \label{eq19}
G_{m}^{ideal\;conductor} \left( {s} \right) = - {\frac{{I_{m} \left( {s}
\right)}}{{K_{m} \left( {s} \right)}}}
\end{equation}

\noindent and main contribution to far field is due to $m=0$ term. As a
result the potential of reflected field for $\rho $-orientation of dipole
decreases at infinity more slowly than (\ref{eq13}). It means that dipole
momentum induced in an ideally conducting nanowire tends to infinity when
wire radius goes to zero. Respectively, the decay rate tends to infinity
too. This fact is in agreement with the exact solution of the
electrodynamic problem \cite{Klimov1}, where it was shown that the decay
rate of radially oriented dipole tends to infinity when cylinder radius
tends to zero.

Up to now we considered  the radiative decay rates only. However even for
dielectric nanobodies the nonradiative losses can be substantial. Here the
nonradiative losses are due to losses inside nanobody, which are proportional
to imaginary part of dielectric permittivity (or conductivity). To take the
nonradiative losses into account, one should use $\left( \ref{eq14_1}\right)$.

In our case, the quasistatic solution ${\rm {\bf E}}^{\left( {R}\right) }$
is given by

\begin{equation}
{\rm {\bf E}}^{\left( {R}\right) }=-\nabla \left( {{\rm {\bf d}}_{0}{\nabla }%
^{\prime }}\right) \tilde{\varphi}^{\left( {2}\right) }\left( {{\rm {\bf r}},%
{\rm {\bf {r}^{\prime }}}}\right) ,  \label{eq22}
\end{equation}

\noindent where $\tilde{\varphi}^{\left( {2}\right) }\left( {{\rm {\bf r}},%
{\rm {\bf {r}^{\prime }}}}\right) $ is given by (\ref{eq9}). Substituting
this expression into $\left( \ref{eq14_1}\right)$ we obtain the expressions 
for the nonradiative decay rates near a nanofiber:

\begin{equation}
\begin{array}{l}
\displaystyle\left( {{\frac{{\gamma }^{nonradiative}}{{\gamma _{0}}}}}%
\right) _{\rho }=-{\frac{{3}}{{\pi k^{3}}}}{\sum\limits_{m=0}^{\infty }{%
\left( {2-\delta _{m,0}}\right) }}{\int\limits_{0}^{\infty }}%
dhh^{2}K_{m}^{\prime 2}\left( h{{\rho }^{\prime }}\right) Im\left( {%
G_{m}\left( ha\right) }\right) \\ 
\displaystyle\left( {{\frac{{\gamma }^{nonradiative}}{{\gamma _{0}}}}}%
\right) _{\varphi }=-{\frac{{6}}{{\pi k^{3}}}}{\sum\limits_{m=0}^{\infty }{{%
\frac{{m^{2}}}{{\rho }^{\prime 2}}}{\int\limits_{0}^{\infty }}}}dhK{{%
_{m}^{2}\left( h{{\rho }^{\prime }}\right) }}Im\left( {G_{m}\left( ha\right) 
}\right) \\ 
\displaystyle\left( {{\frac{{\gamma }^{nonradiative}}{{\gamma _{0}}}}}%
\right) _{z}=-{\frac{{3}}{{\pi k^{3}}}}{\sum\limits_{m=0}^{\infty }{\left( {%
2-\delta _{m,0}}\right) }}{\int\limits_{0}^{\infty }}dhh^{2}K_{m}^{2}\left( h%
{{\rho }^{\prime }}\right) Im\left( {G_{m}\left( ha\right) }\right)
\end{array}
\label{eq23}
\end{equation}

If an atom is placed in close vicinity to surface of a nanofiber $\left( {{%
\rho} ^{\prime}\to a} \right)$, the above  expressions can be reduced to

\begin{equation}
\begin{array}{l}
\displaystyle\left( {{\frac{{\gamma }^{nonradiative}}{{\gamma _{0}}}}}%
\right) _{\rho }=Im\left( {{\frac{{\varepsilon -1}}{{\varepsilon +1}}}}%
\right) {\frac{{3}}{{16k^{3}\left( {{\rho }^{\prime }-a}\right) ^{3}}}} \\ 
\displaystyle\left( {{\frac{{\gamma }^{nonradiative}}{{\gamma _{0}}}}}%
\right) _{\varphi }=\left( {{\frac{{\gamma^{nonradiative}}}{{\gamma _{0}}}}}\right)
_{z}=Im\left( {{\frac{{\varepsilon -1}}{{\varepsilon +1}}}}\right) {\frac{{3}%
}{{32k^{3}\left( {{\rho }^{\prime }-a}\right) ^{3}}}}
\end{array}
\label{eq24}
\end{equation}

It should be noted, these expressions are similar to the case of an atom
near plane dielectric interface, where the reflected field can be described
by image dipoles. Comparing (\ref{eq24}) and (\ref{eq16}) one can see that
for usual dielectrics with low losses (for fused silica $Im\varepsilon \sim
10^{-9})$ the nonradiative losses are very small for any reasonable distance of
an atom from a surface. However, when the atom is placed very close to the surface,
the nonradiative losses can be enhanced substantially.

\section{Full electrodynamic approach to decay rates near dielectric fiber.}

In previous Section the decay rates were found within quasistatic
approximation without taking into account guided modes. In the present Section,
within full Maxwell's propagation theory, we will find the exact expressions
for decay rates, which include the contributions from guided modes.

According to \cite{Wylie} - \cite{Welsch5}, classical and
quantum-electrodynamic calculations give the same results for dipole
transition rate normalized to its vacuum value. In present section we shall
investigate the influence of a dielectric cylinder on transition rates
within classical approach, where the full decay rate can be expressed
through classical reflected field at the atom position (\ref{eq14_1}). Thus
the reflected field must be found to determine the total decay rate. To find
the reflected field, it is necessary to solve the full system of Maxwell's
equations where dipole momentum of oscillator is a source and to use
appropriate boundary conditions.

To find the reflected field we follow an approach develoded by Katsenelenbaum \cite
{Kats1,Kats2} and Wait \cite{Wait}. According to that approach one
should expand all fields over cylinder harmonics. For longitudinal
components we have the following expressions: 
\begin{equation}
\begin{array}{l}
E_{z}^{\left( R\right) }={\sum\limits_{m=-\infty }^{\infty } {\int {dh\;e^{im\varphi +ihz}}}}%
E_{z,mh}^{\left( R\right) }\left( {\rho }\right) ={\sum\limits_{m=-\infty }^{\infty } {\int {%
dh\;e^{im\varphi +ihz}H_{m}^{\left( {1}\right) }\left( {\nu _{2}\rho }%
\right) }}}a_{mh} \\ 
E_{0z}={\sum\limits_{m=-\infty }^{\infty } {\int {dh\;e^{im\varphi +ihz}}}}E_{0z,mh}\left( {\rho 
}\right) \\ 
E_{z}^{\left( T\right) }={\sum\limits_{m=-\infty }^{\infty } {\int {dh\;e^{im\varphi +ihz}}}}%
E_{z,mh}^{\left( T\right) }\left( {\rho }\right) ={\sum\limits_{m=-\infty }^{\infty } {\int {%
dh\;e^{im\varphi +ihz}J_{m}\left( {\nu _{1}\rho }\right) c_{mh}}}} \\ 
B_{z}^{\left( R\right) }={\sum\limits_{m=-\infty }^{\infty } {\int {dh\;e^{im\varphi +ihz}}}}%
B_{z,mh}^{\left( R\right) }\left( {\rho }\right) ={\sum\limits_{m=-\infty }^{\infty } {\int {%
dh\;e^{im\varphi +ihz}H_{m}^{\left( {1}\right) }\left( {\nu _{2}\rho }%
\right) }}}b_{mh} \\ 
B_{0z}={\sum\limits_{m=-\infty }^{\infty } {\int {dh\;e^{im\varphi +ihz}}}}B_{0z,mh}\left( {\rho 
}\right) \\ 
B_{z}^{\left( T\right) }={\sum\limits_{m=-\infty }^{\infty } {\int {dh\;e^{im\varphi +ihz}}}}%
B_{z,mh}^{\left( T\right) }\left( {\rho }\right) ={\sum\limits_{m=-\infty }^{\infty } {\int {%
dh\;e^{im\varphi +ihz}J_{m}\left( {\nu _{1}\rho }\right) d_{mh}}}}
\end{array}
\label{eq25}
\end{equation}

Here superscripts {\it \ }$R$ and $T${\it \ }and subscript{\it \ }$0$
correspond to dipole field in free space, reflected and transmitted inside
cylinder fields, respectively, and $\nu _{1}=\sqrt{\varepsilon k^{2}-h^{2}}$
\ and $\nu _{2}=\sqrt{k^{2}-h^{2}}$ are the radial wavenumbers inside and
outside fiber. The $a_{mh},b_{mh},c_{mh},d_{mh}$ coefficients are to be
determined. We choose the branch cut such as $Im\left( {\nu _{1}}\right)
,Im\left( {\nu _{2}}\right) >0$ in complex plane of longitudinal wavenumber $%
h$. To ensure the decreasing of fields at space infinity $\left( {\rho
\rightarrow \infty }\right) $ the integration in (\ref{eq25}) should be over
path $C_{{\rm 1}}$, which is shown in Fig.2a.

The rest of the field components ($\rho $,$\varphi )$ can be expressed
through z-components of electric and magnetic fields:

\begin{equation}
\begin{array}{l}
E_{\rho ,mh}={\frac{{ih}}{{\nu ^{2}}}}{\frac{{\partial E_{z,mh}}}{{\partial
\rho }}}-{\frac{{km}}{{\rho \nu ^{2}}}}B_{z,mh}; \\ 
\\ 
E_{\varphi ,mh}=-{\frac{{hm}}{{\rho \nu ^{2}}}}E_{z,mh}-{\frac{{ik}}{{\nu
^{2}}}}{\frac{{\partial B_{z,mh}}}{{\partial \rho }}} \\ 
\\ 
B_{\rho ,mh}={\frac{{k\varepsilon m}}{{\rho \nu ^{2}}}}E_{z,mh}+{\frac{{ih}}{%
{\nu ^{2}}}}{\frac{{\partial B_{z,mh}}}{{\partial \rho }}} \\ 
\\ 
B_{\varphi ,mh}={\frac{{ik\varepsilon }}{{\nu ^{2}}}}{\frac{{\partial
E_{z,mh}}}{{\partial \rho }}}-{\frac{{mh}}{{\rho \nu ^{2}}}}B_{z,mh};
\end{array}
\label{eq26}
\end{equation}

In (\ref{eq25}),(\ref{eq26}) the subscripts $m$ and $h$ denote the
appropriate Fourier transformation over $\varphi $ and z, and $\nu =\nu _{1}$
or $\nu _{2}$ for corresponding space region.

For free field one can obtain the following expressions \cite{Jackson}:

\begin{equation}
\begin{array}{l}
\displaystyle{\rm {\bf B}}=rot{\rm {\bf A}},\quad {\rm {\bf E}}={\frac{{i}}{{%
k}}}rot{\rm {\bf H}} \\ 
\displaystyle{\rm {\bf A}}=-ik{\rm {\bf d}}_{0}{\frac{{e^{ik{\left| {{\rm 
{\bf r}}-{\rm {\bf {r}^{\prime }}}}\right| }}}}{{{\left| {{\rm {\bf r}}-{\rm 
{\bf {r}^{\prime }}}}\right| }}}}
\end{array}
\label{eq27}
\end{equation}

\noindent where ${\rm {\bf r}}=\left( {\rho ,\varphi ,z}\right) $ and ${\rm 
{\bf {r}^{\prime }}}=\left( {{\rho }^{\prime },{\varphi }^{\prime },{z}%
^{\prime }}\right) $ are radius vectors of observation point and atom
position.

Using the well known expression \cite{Markov}$ 
\left(\rho <\rho ^{\prime}\right) $

\begin{equation}
{\frac{{e^{ik{\left| {{\rm {\bf r}}-{\rm {\bf {r}^{\prime }}}}\right| }}}}{{{%
\left| {{\rm {\bf r}}-{\rm {\bf {r}^{\prime }}}}\right| }}}}={\frac{{i}}{{2}}%
}{\sum\limits_{m=-\infty }^{\infty } {{\oint\limits_{C_{1}}{dh\;e^{im\left( {\varphi -{\varphi }%
^{\prime }}\right) +ih\left( {z-{z}^{\prime }}\right) }J_{m}\left( {\nu
_{2}\rho }\right) H_{m}^{\left( {1}\right) }\left( {\nu _{2}\rho }^{\prime
}\right) }}}}  \label{eq28}
\end{equation}

\noindent where integration part should be as shown in Fig.2a, one can find
the expressions for longitudinal components of the free dipole  fields near
the fiber surface. For $\rho $ oriented dipole in the region between fiber
and dipole $\left( {a\leq \rho <\rho }^{\prime }\right) $ we have

\begin{equation}
\begin{array}{l}
B_{0z,mh}=-{\frac{{id_{0,\rho }km}}{{2{\rho }^{\prime }}}}J_{m}\left( {\nu
_{2}\rho }\right) H_{m}^{\left( {1}\right) }\left( {\nu _{2}{\rho }^{\prime }%
}\right) \\ 
\\ 
E_{0z,mh}={\frac{{d_{0,\rho }h\nu _{2}}}{{2}}}J_{m}\left( {\nu _{2}\rho }%
\right) {{{\frac{{d}}{{dz}}}H_{m}^{\left( {1}\right) }\left( {z}\right) }}%
_{z=\nu _{2}{\rho }^{\prime }} \\ 
\\ 
E_{0\varphi ,mh}=-{\frac{{md_{0,\rho }}}{{2\nu _{2}{\rho }^{\prime }}}}%
\left( {k^{2}{{{\frac{{d}}{{dz}}}J_{m}\left( {z}\right) }}_{z=\nu _{2}\rho
}H_{m}^{\left( {1}\right) }\left( {\nu _{2}{\rho }^{\prime }}\right) +h^{2}{%
\frac{{{\rho }^{\prime }}}{{\rho }}}J_{m}\left( {\nu _{2}\rho }\right) {{{%
\frac{{d}}{{dz}}}H_{m}^{\left( {1}\right) }\left( {z}\right) }}_{z=\nu _{2}{%
\rho }^{\prime }}}\right) \\ 
\\ 
B_{0\varphi ,mh}={\frac{{ihkd_{0,\rho }}}{{2}}}\left( {{\frac{{m^{2}}}{{\nu
_{2}^{2}{\rho }^{\prime }\rho }}}J_{m}\left( {\nu _{2}\rho }\right)
H_{m}^{\left( {1}\right) }\left( {\nu _{2}{\rho }^{\prime }}\right) +{{{%
\frac{{d}}{{dz}}}J_{m}\left( {z}\right) }}_{z=\nu _{2}\rho _{{}}}{{{\frac{{d}%
}{{dz}}}H_{m}^{\left( {1}\right) }\left( {z}\right) }}_{z=\nu _{2}{\rho }%
^{\prime }}}\right)
\end{array}
\label{eq29}
\end{equation}

\noindent while for $z$ oriented dipole we have

\begin{equation}
\begin{array}{l}
B_{0z,mh}=0 \\ 
E_{0z,mh}={\frac{{id_{0,z}\nu _{2}^{2}}}{{2}}}J_{m}\left( {\nu _{2}\rho }%
\right) H_{m}^{\left( {1}\right) }\left( {\nu _{2}{\rho }^{\prime }}\right)
\\ 
\\ 
E_{0\varphi ,mh}=-{\frac{{imhd_{0,z}}}{{2\rho }}}J_{m}\left( {\nu _{2}\rho }%
\right) H_{m}^{\left( {1}\right) }\left( {\nu _{2}{\rho }^{\prime }}\right)
\\ 
\\ 
B_{0\varphi ,mh}=-{\frac{{d_{0,z}k\nu _{2}}}{{2}}}{\left. {{\frac{{d}}{{dz}}}%
J_{m}\left( {z}\right) }\right| }_{z=\nu _{2}\rho }H_{m}^{\left( {1}\right)
}\left( {\nu _{2}{\rho }^{\prime }}\right)
\end{array}
\label{eq30}
\end{equation}

Finally, for $\varphi $ orientation of dipole momentum we have the following
expressions

\begin{equation}
\begin{array}{l}
B_{0z,mh}=-{\frac{{d_{0,\varphi }k\nu _{2}}}{{2}}}J_{m}\left( {\nu _{2}\rho }%
\right) {{{\frac{{d}}{{dz}}}H_{m}^{\left( {1}\right) }\left( {z}\right) }}%
_{z=\nu _{2}{\rho }^{\prime }} \\ 
\\ 
B_{0\varphi ,mh}={\frac{{hkmd_{0,\varphi }}}{{2\nu _{2}\rho }}}\left( {%
J_{m}\left( {\nu _{2}\rho }\right) {{{\frac{{d}}{{dz}}}H_{m}^{\left( {1}%
\right) }\left( {z}\right) }}_{z=\nu _{2}{\rho }^{\prime }}+{\frac{{\rho }}{{%
{\rho }^{\prime }}}}{\left. {H_{m}^{\left( {1}\right) }\left( {\nu _{2}{\rho 
}^{\prime }}\right) {\frac{{d}}{{dz}}}J_{m}\left( {z}\right) }\right| }%
_{z=\nu _{2}\rho _{{}}}}\right) \\ 
\\ 
E_{0z,mh}=-{\frac{{id_{0,\varphi }hm}}{{2{\rho }^{\prime }}}}J_{m}\left( {%
\nu _{2}\rho }\right) H_{m}^{\left( {1}\right) }\left( {\nu _{2}{\rho }%
^{\prime }}\right) \\ 
\\ 
E_{0\varphi ,mh}={\frac{{id_{0,\varphi }}}{{2}}}\left( {k^{2}{{{\frac{{d}}{{%
dz}}}J_{m}\left( {z}\right) }}_{z=\nu _{2}\rho }{{{\frac{{d}}{{dz}}}%
H_{m}^{\left( {1}\right) }\left( {z}\right) }}_{z=\nu _{2}{\rho }^{\prime }}+%
{\frac{{h^{2}m^{2}}}{{\left( {\nu _{2}{\rho }^{\prime }}\right) \left( {\nu
_{2}\rho }\right) }}}J_{m}\left( {\nu _{2}\rho }\right) H_{m}^{\left( {1}%
\right) }\left( {\nu _{2}{\rho }^{\prime }}\right) }\right)
\end{array}
\label{eq31}
\end{equation}

To find coefficients in transmitted and reflected fields one should take
into account boundary conditions on the surface of dielectric cylinder. As a
result we have a system of 4 equations for 4 unknown coefficients.

\begin{equation}
\begin{array}{l}
H_{m}^{\left( {1}\right) }\left( {z_{2}}\right) a_{mh}-J_{m}\left( {z_{1}}%
\right) c_{mh}=-E_{0z,mh}\left( {\rho =a}\right)  \\ 
\\ 
H_{m}^{\left( {1}\right) }\left( {z_{2}}\right) b_{mh}-J_{m}\left( {z_{1}}%
\right) d_{mh}=-H_{0z,mh}\left( {\rho =a}\right)  \\ 
\\ 
{\frac{{mh}}{{\nu _{2}^{2}a}}}H_{m}^{\left( {1}\right) }\left( {z_{2}}%
\right) a_{mh}+{\frac{{ik}}{{\nu _{2}}}}{\frac{{d}}{{dz_{2}}}}H_{m}^{\left( {%
1}\right) }\left( {z_{2}}\right) b_{mh}-{\frac{{mh}}{{\nu _{1}^{2}a}}}%
J_{m}\left( {z_{1}}\right) c_{mh}-{\frac{{ik}}{{\nu _{1}}}}{\frac{{d}}{{%
dz_{1}}}}J_{m}\left( {z_{1}}\right) d_{mh}=E_{0\varphi ,mh}\left( {\rho =a}%
\right)  \\ 
\\ 
{\frac{{ik}}{{\nu _{2}}}}{\frac{{d}}{{dz_{2}}}}H_{m}^{\left( {1}\right)
}\left( {z_{2}}\right) a_{mh}-{\frac{{mh}}{{\nu _{2}^{2}a}}}H_{m}^{\left( {1}%
\right) }\left( {z_{2}}\right) b_{mh}-{\frac{{ik\varepsilon }}{{\nu _{1}}}}{%
\frac{{d}}{{dz_{1}}}}J_{m}\left( {z_{1}}\right) c_{mh}+{\frac{{mh}}{{\nu
_{1}^{2}a}}}J_{m}\left( {z_{1}}\right) d_{mh}=-H_{0\varphi ,mh}\left( {\rho
=a}\right) 
\end{array}
\label{eq32}
\end{equation}

\noindent where we use abbreviation $z_{1,2}=\nu _{1,2}a$.

The reflected electric fields are determined only by $a_{mh}$ and $b_{mh}$
coefficients, which can be simplified to

\begin{equation}
a_{mh}={\frac{{na}}{{P^{2}+QR}}},\quad b_{mh}={\frac{{nb}}{{P^{2}+QR}}}
\label{eq33}
\end{equation}

\noindent where

\begin{equation}
\begin{array}{l}
na=\nu _{1}^{2}\nu _{2}^{2}aJ_{m}\left( {z_{1}}\right) PE_{0\varphi ,mh}+\nu
_{2}^{2}\left( {J_{m}\left( {z_{1}}\right) }hmP{+}ka{\varepsilon \nu _{1}{%
\frac{{d}}{{dz_{1}}}}J_{m}\left( {z_{1}}\right) Q}\right) E_{0z,mh} \\ 
\\ 
+i\nu _{1}^{2}\nu _{2}^{2}aJ_{m}\left( {z_{1}}\right) QB_{0\varphi
,mh}-imh\nu _{1}\nu _{2}J_{m}\left( {z_{1}}\right) SB_{0z,mh}
\end{array}
\label{eq34}
\end{equation}

\begin{equation}
\begin{array}{l}
nb=\nu _{1}^{2}\nu _{2}^{2}aJ_{m}\left( {z_{1}}\right) PB_{0\varphi ,mh}+\nu
_{2}^{2}\left( {J_{m}\left( {z_{1}}\right) }hmP{-}ak{\nu _{1}{\frac{{d}}{{%
dz_{1}}}}J_{m}\left( {z_{1}}\right) R}\right) B_{0z,mh} \\ 
\\ 
+i\nu _{1}^{2}\nu _{2}^{2}aJ_{m}\left( {z_{1}}\right) RE_{0\varphi
,mh}+imh\nu _{1}\nu _{2}J_{m}\left( {z_{1}}\right) TE_{0z,mh}
\end{array}
\label{eq35}
\end{equation}

\noindent and

\begin{equation}  \label{eq36}
\begin{array}{l}
P = hmk^{2}J_{m} \left( {z_{1}} \right)H_{m}^{\left( {1} \right)} \left( {%
z_{2}} \right)\left( {\varepsilon - 1} \right) \\ 
\\ 
Q = - \nu _{1} \nu _{2} ak\left( {\nu _{1} J_{m} \left( {z_{1}} \right){%
\frac{{d}}{{dz_{2}}} }H_{m}^{\left( {1} \right)} \left( {z_{2}} \right) -
\nu _{2} H_{m}^{\left( {1} \right)} \left( {z_{2}} \right){\frac{{d}}{{dz_{1}%
}} }J_{m} \left( {z_{1}} \right)} \right) \\ 
\\ 
R = \nu _{1} \nu _{2} ak\left( {\nu _{1} J_{m} \left( {z_{1}} \right){\frac{{%
d}}{{dz_{2}}} }H_{m}^{\left( {1} \right)} \left( {z_{2}} \right) - \nu _{2}
\varepsilon H_{m}^{\left( {1} \right)} \left( {z_{2}} \right){\frac{{d}}{{%
dz_{1}}} }J_{m} \left( {z_{1}} \right)} \right) \\ 
\\ 
S = \nu _{1} \nu _{2} ak\left( {\nu _{2} J_{m} \left( {z_{1}} \right){\frac{{%
d}}{{dz_{2}}} }H_{m}^{\left( {1} \right)} \left( {z_{2}} \right) - \nu _{1}
H_{m}^{\left( {1} \right)} \left( {z_{2}} \right){\frac{{d}}{{dz_{1}}} }%
J_{m} \left( {z_{1}} \right)} \right) \\ 
\\ 
T = \nu _{1} \nu _{2} ak\left( {\nu _{2} J_{m} \left( {z_{1}} \right){\frac{{%
d}}{{dz_{2}}} }H_{m}^{\left( {1} \right)} \left( {z_{2}} \right) - \nu _{1}
\varepsilon H_{m}^{\left( {1} \right)} \left( {z_{2}} \right){\frac{{d}}{{%
dz_{1}}} }J_{m} \left( {z_{1}} \right)} \right)
\end{array}
\end{equation}

By substituting the expressions for reflected field (\ref{eq25})- (\ref{eq26}%
) into general expression (\ref{eq10_1}) we obtain final expressions for total
decay rates:

\begin{equation}
\left( {{\frac{{\gamma }^{total}}{{\gamma _{0}}}}}\right) _{\rho }=1+{\frac{{%
3}}{{2}}}Im{\frac{{{\sum\limits_{m=-\infty }^{\infty } {{\oint\limits_{C_{1}}{dh{\frac{{1}}{{\nu
_{2}^{2}}}}{\left[ {ih\nu _{2}{\frac{{d}}{{dz}}}H_{m}^{\left( {1}\right)
}\left( {z}\right) a_{mh}-{\frac{{km}}{{{\rho }^{\prime }}}}H_{m}^{\left( {1}%
\right) }\left( {z}\right) b_{mh}}\right] }_{z=\nu _{2}{\rho }^{\prime }}}}}}%
}}{{d_{0,\rho }k^{3}}}}  \label{eq37}
\end{equation}

\begin{equation}
\left( {{\frac{{\gamma }^{total}}{{\gamma _{0}}}}}\right) _{\varphi }=1+{%
\frac{{3}}{{2}}}Im{\frac{{{\sum\limits_{m=-\infty }^{\infty } {{\oint\limits_{C_{1}}{dh{\frac{{1}%
}{{\nu _{2}^{2}}}}{\left[ {\ -ik\nu _{2}{\frac{{d}}{{dz}}}H_{m}^{\left( {1}%
\right) }\left( {z}\right) b_{mh}-{\frac{{hm}}{{{\rho }^{\prime }}}}%
H_{m}^{\left( {1}\right) }\left( {z}\right) a_{mh}}\right] }_{z=\nu _{2}{%
\rho }^{\prime }}}}}}}}{{d_{0,\varphi }k^{3}}}}  \label{eq38}
\end{equation}

\begin{equation}
\left( {{\frac{{\gamma }^{total}}{{\gamma _{0}}}}}\right) _{z}=1+{\frac{{3}}{%
{2}}}Im{\frac{{{\sum\limits_{m=-\infty }^{\infty } {{\oint\limits_{C_{1}}{dhH_{m}^{\left( {1}%
\right) }\left( {\nu _{2}{\rho }^{\prime }}\right) a_{mh}}}}}}}{{d_{0,z}k^{3}%
}}}  \label{eq39}
\end{equation}

As  was mentioned above, the important feature of dielectric fiber is the
presence of guided modes, which differs substantially from free space
spherical (or cylindrical) waves. Such modes are effectively used in optical
communications lines. Note that there are no such modes in the case of dielectric
sphere or ideally conducting cylinder. So-called whispering gallery modes,
which occur in the case of sphere or cylinder, are the decaying ones even in the
case of lossless materials.

Thus there are two types of modes in presence of dielectric fiber: free
space radiation modes and waveguided modes. From mathematical point of
view these modes correspond to different types of a spectrum. The guided modes
correspond to discrete spectrum while radiating modes correspond to
continuous part of the spectrum. In this connection the expressions (\ref{eq37}%
)-(\ref{eq39}) are not fully suitable for further analysis, because the
guided modes are not separated here.

The appearance of guided modes is connected with poles in subintegral
expressions in (\ref{eq37})- (\ref{eq39}). One can show that in the case of
a lossless dielectric these poles are situated on the real axis of $h$ between 
$k$ and $k\sqrt{\varepsilon }$ and the number of poles is finite\cite{Kats1,Kats2}.

Now transforming integration contour  from $C_{1}$ to $C_{2}$, as shown
in Fig.2b, and applying the residue theorem, one can  separate
radiating and guided modes

\begin{equation}
\begin{array}{l}
\left( {{\frac{{\gamma }^{total}}{{\gamma _{0}}}}}\right) _{\rho }=1+{\frac{{%
3}}{{2}}}{\rm Im}{\frac{{{\sum\limits_{m=-\infty }^{\infty } {{\oint\limits_{C_{2}}{dh{\frac{{1}%
}{{\nu _{2}^{2}}}}{\left[ {ih\nu _{2}{\frac{{d}}{{dz}}}H_{m}^{\left( {1}%
\right) }\left( {z}\right) a_{mh}-{\frac{{km}}{{{\rho }^{\prime }}}}%
H_{m}^{\left( {1}\right) }\left( {z}\right) b_{mh}}\right] }_{z=\nu _{2}{%
\rho }^{\prime }}}}}}}}{{d_{0,\rho }^{{}}k^{3}}}} \\ 
\\ 
+3\pi {\rm Re}{\sum\limits_{m=-\infty }^{\infty } {{\sum\limits_{h_{\alpha },m}{{\rm R}{\rm e}%
{\rm s}{\left[ {{\frac{{{\frac{{1}}{{\nu _{2}^{2}}}}{\left[ {ih\nu _{2}{%
\frac{{d}}{{dz}}}H_{m}^{\left( {1}\right) }\left( {z}\right) a_{mh}-{\frac{{%
km}}{{{\rho }^{\prime }}}}H_{m}^{\left( {1}\right) }\left( {z}\right) b_{mh}}%
\right] }_{z=\nu _{2}{\rho }^{\prime }}}}{{d_{0,\rho }k^{3}}}}}\right] }}}}}%
_{h=h_{a,m}}
\end{array}
\label{eq40}
\end{equation}

\begin{equation}
\begin{array}{l}
\left( {{\frac{{\gamma }^{total}}{{\gamma _{0}}}}}\right) _{\varphi }=1+{%
\frac{{3}}{{2}}}{\rm Im}{\frac{{{\sum\limits_{m=-\infty }^{\infty } {{\oint\limits_{C_{2}}{dh{%
\frac{{1}}{{\nu _{2}^{2}}}}{\left[ {\ -ik\nu _{2}{\frac{{d}}{{dz}}}%
H_{m}^{\left( {1}\right) }\left( {z}\right) b_{mh}-{\frac{{hm}}{{{\rho }%
^{\prime }}}}H_{m}^{\left( {1}\right) }\left( {z}\right) a_{mh}}\right] }%
_{z=\nu _{2}{\rho }^{\prime }}}}}}}}{{d_{0,\varphi }k^{3}}}} \\ 
\\ 
+3\pi {\rm Re}{\sum\limits_{m=-\infty }^{\infty } {{\sum\limits_{h_{\alpha },m}{{\rm R}{\rm e}%
{\rm s}{\left[ {{\frac{{{\frac{{1}}{{\nu _{2}^{2}}}}{\left[ {\ -ik\nu _{2}{%
\frac{{d}}{{dz}}}H_{m}^{\left( {1}\right) }\left( {z}\right) b_{mh}-{\frac{{%
hm}}{{{\rho }^{\prime }}}}H_{m}^{\left( {1}\right) }\left( {z}\right) a_{mh}}%
\right] }_{z=\nu _{2}{\rho }^{\prime }}}}{{d_{0,\varphi }k^{3}}}}}\right] }}}%
}}_{h=h_{a,m}}
\end{array}
\label{eq41}
\end{equation}

\begin{equation}
\begin{array}{l}
\left( {{\frac{{\gamma }^{total}}{{\gamma _{0}}}}}\right) _{z}=1+{\frac{{3}}{%
{2}}}{\rm Im}{\frac{{{\sum\limits_{m=-\infty }^{\infty } {{\oint\limits_{C_{2}}{dh{\left[ {%
H_{m}^{\left( {1}\right) }\left( \nu _{2}\rho ^{\prime }\right) a_{mh}}%
\right] }}}}}}}{{d_{0,z}k^{3}}}} \\ 
\\ 
+3\pi {\rm Re}{\sum\limits_{m=-\infty }^{\infty } {{\sum\limits_{h_{\alpha ,m}}{{\rm R}{\rm e}%
{\rm s}{\left[ {{\frac{{{\left[ {H_{m}^{\left( {1}\right) }\left( \nu
_{2}\rho ^{\prime }\right) a_{mh}}\right] }}}{{d_{0,z}k^{3}}}}}\right] }}}}}%
_{h=h_{a,m}}
\end{array}
\label{eq42}
\end{equation}

\noindent where the sum is over all poles $h_{\alpha ,m}$ of subintegral
functions , {\it Res} means residue, and $a_{mh},b_{mh}$ are described by (%
\ref{eq33})-(\ref{eq36}) .

In expressions (\ref{eq40})-(\ref{eq42}) the first term corresponds to
spherical waves running to infinity and nonradiative losses, while the
second term corresponds to guided modes. It should be emphasized that in the
case of lossless media the vertical part of branch cut (along imaginary axis
of $h)$ gives no contribution to decay rates. For lossy media this part is
very important because it contributes to nonradiative losses. The
nonradiative decay rates found in previous section (expressions (\ref{eq23}%
)) are the asymptotics of integral over the vertical part of branch cut when 
$ka$ moves to zero.

As  was mentioned above, in the dielectric case there is only a finite
number of guided modes with longitudinal wavevectors between $k$ and $k%
\sqrt {\varepsilon} $. Moreover, for small enough fiber (for nanofiber!) with

\begin{equation}
ka<{\frac{{j}_{0,1}}{\sqrt{\varepsilon -1}},}  \label{eq43}
\end{equation}

\noindent (where\ $j_{0,1}\approx 2.4048$ is the first root of $J_{0}\left(
z\right) $) \ {}the only one guided mode with $m=\pm 1$ exists. Sometimes
such modes are referred to as main or principal modes.

The dependence of longitudinal wavenumber $h$ on cylinder radius or
frequency is determined by dispersion equation

\begin{equation}
D=P^{2}+QR=0,\quad m=\pm 1  \label{eq44}
\end{equation}

\noindent where P , Q and R are defined by (\ref{eq36}).

In the case of nanofibers, $ka << 1$, the asymptotic solution of (\ref{eq44}%
) can be presented in the form:

\begin{equation}
\left( {\frac{{h}}{{k}}}\right) _{10}=1+{\frac{{2}}{{\left( {ka}\right) ^{2}}%
}}\exp \left( {-{\frac{{2}}{{\left( {ka}\right) ^{2}}}}{\frac{{\varepsilon +1%
}}{{\varepsilon -1}}}+{\frac{{\varepsilon +1}}{{4}}}-2\gamma +...}\right) 
\label{eq45}
\end{equation}

\noindent where $\gamma =0.5776$ is the Euler constant.

The exact and asymptotic solutions of (\ref{eq44}) are shown in Fig.3 , where
one can see that asymptotics (\ref{eq45}) presents solution of (\ref{eq44})
correctly if $ka\leq 0.8$. In what follows we restrict ourselves to the case
of nanofiber, where condition (\ref{eq43}) holds true.

To calculate decay rates into guided modes one should know  residues of
the corresponding expression. The residues can be found if one knows the
asymptotic behavior of resonant denominator near pole. In the case of
nanofiber the denominator can be approximated by

\begin{equation}
D=P^{2}+QR\approx k^{6}{\frac{{\left( {\varepsilon -1}\right) ^{3}}}{{\pi
^{2}}}}\left( {ka}\right) ^{2}\left( {\left( h/k\right) /\left( {h/k}\right)
_{10}-1}\right)  \label{eq46}
\end{equation}

\noindent where $\left( {{\frac{{h}}{{k}}}}\right) _{10}$ is given by (\ref
{eq45}).

Let us stress once more that the expressions (\ref{eq40}) - (\ref{eq42}), (%
\ref{eq45}) and  (\ref{eq46}) are valid in the case of any complex dielectric
permittivity. In the case of metallic cylinder, that is, in the case where $%
Re\left( {\varepsilon }\right) < -1$, the poles of subintegral function are
also near the real axis of $h$. But now they correspond to symmetric $(m=0)$
guided modes. More detailed investigation of influence of symmetric $(m=0)$
guided modes on decay rate will be presented in a separate publication \cite
{Klimov4}.

\section{Decay rates near surface of nanofiber without losses.}

The expressions (\ref{eq40}) - (\ref{eq42}) fully describe the problem of
spontaneous emission of an atom placed near a cylinder made of any material.
However, those expressions are too complicated to understand real picture of
decay rate. Our goal is to find simple analytical expressions allowing one  to
estimate decay rate near dielectric \ $(\varepsilon >1)$ cylinder with
radius, which is substantially smaller in comparison with wavelength, $ka<<1$.
Moreover, to obtain simple asymptotes we restrict ourselves to the case of
an atom placed at the surface $\left( {{\rho }^{\prime }\rightarrow a}%
\right) $ of a lossless dielectric cylinder.

\subsection{$z$-orientation of dipole}

Substituting the expression for exciting external fields (\ref{eq30}) into
general expression for decay rate (\ref{eq42}) one can represent the decay
rate for z-oriented dipole momentum in the form:

\begin{equation}
\begin{array}{l}
\left( {{\frac{{\gamma }^{total}}{{\gamma _{0}}}}}\right) _{z}=1-{\frac{{3}}{%
{2}}}{\rm Re}{\sum\limits_{m=-\infty }^{\infty } {{\int\limits_{0}^{k}{{\frac{{dh\nu _{2}^{2}}}{{%
k^{3}}}}{\frac{{H_{m}^{\left( {1}\right) }\left( {\nu _{2}{\rho }^{\prime }}%
\right) ^{2}J_{m}\left( {z_{2}}\right) }}{{H_{m}^{\left( {1}\right) }\left( {%
z_{2}}\right) }}}}}}} \\ 
\\ 
-{\frac{{3}}{{\pi }}}{\rm Im}{\sum\limits_{m=-\infty }^{\infty } {{\int\limits_{0}^{k}{{\frac{{%
dh\nu _{1}^{2}\nu _{2}^{2}}}{{k^{2}}}}{\frac{{H_{m}^{\left( {1}\right)
}\left( {\nu _{2}{\rho }^{\prime }}\right) ^{2}J_{m}\left( {z_{1}}\right) Q}%
}{{H_{m}^{\left( {1}\right) }\left( {z_{2}}\right) D}}}}}}} \\ 
\\ 
-3{\rm Re}{\sum\limits_{m=-\infty }^{\infty } {{\sum\limits_{h_{\alpha ,m}}{\ }}}}{\rm Res}{{{{{%
\left[ {{\frac{{\nu _{1}^{2}\nu _{2}^{2}}}{{k^{2}}}}{\frac{{H_{m}^{\left( {1}%
\right) }\left( {\nu _{2}{\rho }^{\prime }}\right) ^{2}J_{m}\left( {z_{1}}%
\right) Q}}{{H_{m}^{\left( {1}\right) }\left( {z_{2}}\right) D}}}}\right] }}}%
}}_{h_{\alpha ,m}}
\end{array}
\label{eq47}
\end{equation}

\noindent where Q and D are defined by (\ref{eq36}) and (\ref{eq44}), and $%
z_{1,2}=\nu _{1,2}a$.

Note, the first line of (\ref{eq47}) coincides with decay rate near ideally
conducting cylinder \cite{Klimov1}. This expression is valid for real
dielectric permittivities.

In the most interesting case of an atom near surface of nanofiber $\left( {{%
\rho }^{\prime }=a}\right) $ using identity ${\sum\limits_{m}{%
J_{m}^{2}\left( {z}\right) }}=1$ one can simplify (\ref{eq47}) to a more
compact form

\begin{equation}
\begin{array}{l}
\left( {{\frac{{\gamma^{total} }}{{\gamma _{0}}}}}\right) _{z}=-{\frac{{3}}{{\pi }}}%
{\rm Im}{\sum\limits_{m=-\infty }^{\infty } {{\int\limits_{0}^{k}{{\frac{{dh\nu _{1}^{2}\nu
_{2}^{2}}}{{k^{2}}}}{\frac{{H_{m}^{\left( {1}\right) }\left( {z_{2}}\right)
J_{m}\left( {z_{1}}\right) Q}}{{D}}}}}}} \\ 
\\ 
-3{\rm Re}{\sum\limits_{m=-\infty }^{\infty } {{\sum\limits_{h_{\alpha },_{m}}}}}{\rm Res}\left[ 
{{\frac{{\nu _{1}^{2}\nu _{2}^{2}}}{{k^{2}}}}{\frac{{H_{m}^{\left( {1}%
\right) }\left( {z_{2}}\right) J_{m}\left( {z_{1}}\right) Q}}{{D}}}}\right]
_{h_{\alpha ,m}}
\end{array}
\label{eq48}
\end{equation}

The asymptote of (\ref{eq48}) for $ka \ll 1$ has the following form

\begin{equation}  \label{eq49}
\begin{array}{l}
\left( {{\frac{{\gamma^{total}}} {{\gamma _{0}}} }} \right)_{z} \approx 1 - {\frac{{%
\left( {\varepsilon - 1} \right)}}{{75}}}\left( {60\left( {\gamma + \ln ka}
\right) - 47 - {\frac{{45\varepsilon - 15}}{{\left( {\varepsilon + 1}
\right)^{2}}}}} \right)\left( {ka} \right)^{2} + O\left( {\left( {ka}
\right)^{4}} \right) \\ 
\\ 
+ {\frac{{12}}{{\left( {ka} \right)^{4}}}}\exp \left( {\ - {\frac{{2}}{{%
\left( {ka} \right)^{2}}}}{\frac{{\varepsilon + 1}}{{\varepsilon - 1}}} + {%
\frac{{\varepsilon + 1}}{{4}}} - 2\gamma} \right)
\end{array}
\end{equation}

Here the first line describes the radiative losses while exponentially small
second line describes contribution of the principal guided mode with $m=\pm 1
$.

\subsection{$\protect\varphi$ - oriented dipole}

Let us now consider the case of a dipole having $\varphi $ - orientation of dipole
momentum and being located in close vicinity to the surface of nanofiber (${\rho }%
^{\prime }\rightarrow a)$. In the case of lossless media the general
expression (\ref{eq41}) can be simplified to

\begin{equation}
\begin{array}{l}
\left( {{\frac{{\gamma }^{total}}{{\gamma _{0}}}}}\right) _{\varphi }=1-{%
\frac{{3}}{{2}}}{\rm Re}{\sum\limits_{m=-\infty }^{\infty }{{%
\int\limits_{0}^{k}{{\frac{{dh}}{{k}}}}}}}{\left[ {{\frac{{h^{2}m^{2}}}{{%
k^{2}z_{2}^{2}}}}J\left( {z_{2}}\right) H\left( {z_{2}}\right) +{H}^{\prime
}\left( {z_{2}}\right) {J}^{\prime }\left( {z_{2}}\right) }\right] }- \\ 
\\ 
+{\frac{{3}}{{\pi }}}{\rm Im}{\sum\limits_ {m=-\infty }^{\infty } {{%
\int\limits_{0}^{k}{dh}}}}{\frac{{h^{2}m^{2}\nu _{2}J\left( {z_{1}}\right)
H\left( {z_{2}}\right) \left( {\nu _{2}^{3}J\left( {z_{1}}\right) {H}%
^{\prime }\left( {z_{2}}\right) -\nu _{1}^{3}{J}^{\prime }\left( {z_{1}}%
\right) H\left( {z_{2}}\right) }\right) +\nu _{1}\nu _{2}^{2}ka{H}^{\prime
}\left( {z_{2}}\right) {J}^{\prime }\left( {z_{1}}\right) R}}{{kz_{2}D}}} \\ 
\\ 
+3Re{\sum\limits_{m=-\infty }^{\infty }{{\sum\limits_{h_{\alpha ,m}}{\rm Res}%
{{\left[ {{\frac{{h^{2}m^{2}\nu _{2}J\left( {z_{1}}\right) H\left( {z_{2}}%
\right) \left( {\nu _{2}^{3}J\left( {z_{1}}\right) {H}^{\prime }\left( {z_{2}%
}\right) -\nu _{1}^{3}{J}^{\prime }\left( {z_{1}}\right) H\left( {z_{2}}%
\right) }\right) +\nu _{1}\nu _{2}^{2}ka{H}^{\prime }\left( {z_{2}}\right) {J%
}^{\prime }\left( {z_{1}}\right) R}}{{kz_{2}D}}}}\right] }}}}}_{h_{\alpha
,m}}
\end{array}
\label{eq50}
\end{equation}

\noindent \noindent where R and D are defined by (\ref{eq36}) and (\ref{eq44}). Here for brevity we omit
indices $(m)$ in Bessel and Hankel functions of first kind and use prime to
denote derivative of Hankel and Bessel functions.

Using the identities

\begin{equation}
{\sum\limits_{n=-\infty }^{n=\infty }{J_{n}^{2}\left( {z}\right) =1}};\quad {%
\sum\limits_{n=-\infty }^{n=\infty }{n^{2}J_{n}^{2}\left( {z}\right) ={\frac{%
{z^{2}}}{{2}}}}};\quad {\sum\limits_{n=-\infty }^{n=\infty }{\left( {{\frac{{%
dJ_{n}\left( {z}\right) }}{{dz}}}}\right) ^{2}={\frac{{1}}{{2}}}}}
\label{eq51}
\end{equation}

\noindent one can show that first line in (\ref{eq50}) is equal to zero.
Finally the expression for radiative decay rate of a $\varphi $-oriented
dipole placed at the surface of nanofiber acquires the following form

\begin{equation}
\begin{array}{l}
\left( {{\frac{{\gamma }^{total}}{{\gamma _{0}}}}}\right) _{\varphi }={\frac{%
{3}}{{\pi }}}{\rm Im}{\sum\limits_{m=-\infty }^{\infty }{{\int\limits_{0}^{k}%
{dh}}}}{\frac{{h^{2}m^{2}\nu _{2}J\left( {z_{1}}\right) H\left( {z_{2}}%
\right) \left( {\nu _{2}^{3}J\left( {z_{1}}\right) {H}^{\prime }\left( {z_{2}%
}\right) -\nu _{1}^{3}{J}^{\prime }\left( {z_{1}}\right) H\left( {z_{2}}%
\right) }\right) +\nu _{1}\nu _{2}^{2}ka{H}^{\prime }\left( {z_{2}}\right) {J%
}^{\prime }\left( {z_{1}}\right) R}}{{kz_{2}D}}} \\ 
\\ 
+3{\rm Re}{\sum\limits_{m=-\infty }^{\infty }{{\sum\limits_{h_{\alpha ,m}}{%
{\rm R}{\rm e}{\rm s}{\left[ {{\frac{{h^{2}m^{2}\nu _{2}J\left( {z_{1}}%
\right) H\left( {z_{2}}\right) \left( {\nu _{2}^{3}J\left( {z_{1}}\right) {H}%
^{\prime }\left( {z_{2}}\right) -\nu _{1}^{3}{J}^{\prime }\left( {z_{1}}%
\right) H\left( {z_{2}}\right) }\right) +\nu _{1}\nu _{2}^{2}ka{H}^{\prime
}\left( {z_{2}}\right) {J}^{\prime }\left( {z_{1}}\right) R}}{{kz_{2}D}}}}%
\right] }}}}}_{h_{\alpha ,m}}
\end{array}
\label{eq52}
\end{equation}

The asymptote of (\ref{eq52}) for $ka<<1$ has the following form

\begin{equation}
\begin{array}{l}
\left( {{\frac{{\gamma }^{total}}{{\gamma _{0}}}}}\right) _{\varphi }=\left( 
{{\frac{{2}}{{\varepsilon +1}}}}\right) ^{2}+{\frac{{\left( {\varepsilon -1}%
\right) \left( {75\varepsilon ^{2}+2081-1680\left( {\gamma +\ln \left( {ka}%
\right) }\right) }\right) \left( {ka}\right) ^{2}}}{{300\left( {\varepsilon
+1}\right) ^{3}}}}+O\left( {\left( {ka}\right) ^{4}}\right) \\ 
\\ 
+{\frac{{48}}{{\left( {\varepsilon -1}\right) ^{2}\left( {ka}\right) ^{6}}}}%
\exp \left( {\ -{\frac{{2}}{{\left( {ka}\right) ^{2}}}}{\frac{{\varepsilon +1%
}}{{\varepsilon -1}}}+{\frac{{\varepsilon +1}}{{4}}}-2\gamma }\right)
\end{array}
\label{eq53}
\end{equation}

Here the first line describes the radiative losses while exponentially small
second line describes contribution of the main guided mode  $m=\pm 1$.

\subsection{$\protect\rho $ - oriented dipole}

The case of radially oriented dipole momentum is more complicated for
analysis. So we again restrict ourselves to the most interesting case, when
atom is near  to dielectric surface, that is, we consider ${\rho }^{\prime }=a$
case. In the case of lossless media the general expression (\ref{eq40}) can
be simplified to

\begin{equation}
\begin{array}{l}
\left( {{\frac{{\gamma }^{total}}{{\gamma _{0}}}}}\right) _{\rho }=1-{\frac{{%
3}}{{2}}}Re{\sum\limits_{m=-\infty }^{\infty }{{\int\limits_{0}^{k}{{\frac{{%
dh}}{{k}}}}}}}{{{\frac{{h^{2}}}{{k^{2}}}}{H}}^{\prime 2}{\left( {z_{2}}%
\right) {\frac{{J\left( {z_{2}}\right) }}{{H\left( {z_{2}}\right) }}}+{\frac{%
{m^{2}H\left( {z_{2}}\right) J\left( {z_{2}}\right) }}{{z_{2}^{2}}}}}}- \\ 
\\ 
-{\frac{{2i\nu _{1}^{2}}}{{\pi kz_{2}^{2}}}}{\frac{{J\left( {z_{1}}\right) }%
}{{H\left( {z_{2}}\right) }}}{{{\left\{ {{\frac{{2h^{2}k^{3}m^{2}z_{2}\left( 
{\varepsilon -1}\right) J\left( {z_{1}}\right) H\left( {z_{2}}\right) ^{2}{H}%
^{\prime }\left( {z_{2}}\right) +z_{2}^{2}h^{2}Q{H}^{\prime }{}^{2}\left( {%
z_{2}}\right) -m^{2}k^{2}H\left( {z_{2}}\right) ^{2}R}}{{D}}}}\right\} }}}
\\ 
\\ 
-3Re{\sum\limits_{m=-\infty }^{\infty }{{\sum\limits_{h_{\alpha ,m}}}}}{\rm %
Res}\left[ {{\frac{{\nu _{1}^{2}}}{{k^{2}z_{2}^{2}}}}{\frac{{J\left( {z_{1}}%
\right) }}{{H\left( {z_{2}}\right) }}}{\left\{ {{\frac{{%
2h^{2}k^{3}m^{2}z_{2}\left( {\varepsilon -1}\right) J\left( {z_{1}}\right)
H\left( {z_{2}}\right) ^{2}{H}^{\prime }\left( {z_{2}}\right)
+z_{2}^{2}h^{2}Q{H}^{\prime }{}^{2}\left( {z_{2}}\right) -m^{2}k^{2}H\left( {%
z_{2}}\right) ^{2}R}}{{D}}}}\right\} }}\right] _{h_{\alpha ,m}}
\end{array}
\label{eq54}
\end{equation}

\noindent \noindent where Q,R and D are defined by (\ref{eq36}) and (\ref{eq44}).

In the case $\varepsilon \rightarrow \infty $, from integral over cut (first
two lines in (\ref{eq54})) one can reveal the decay rate for an atom near
an ideally conducting cylinder \cite{Klimov1}:

\begin{equation}
\left( {{\frac{{\gamma }^{total}}{{\gamma _{0}}}}}\right) _{\rho }={\frac{{6}%
}{{\pi ^{2}}}}{\sum\limits_{m=-\infty }^{\infty }{{\int\limits_{0}^{k}{dh{%
\frac{{h^{2}}}{{k^{3}\left( {\nu }_{2}{a}\right) ^{2}}}}}}}}{\frac{{1}}{{{%
\left| {H_{m}^{\left( {1}\right) }\left( {\nu }_{2}{a}\right) }\right| }^{2}}%
}}+{\frac{{6}}{{\pi ^{2}}}}{\sum\limits_{m=-\infty }^{\infty }{{%
\int\limits_{0}^{k}{dh{\frac{{m^{2}}}{{k\left( {\nu }_{2}{a}\right) ^{4}}}}{%
\frac{{1}}{{{\left| {{{{\frac{{d}}{{dz}}}\left( {H_{m}^{\left( {1}\right)
}\left( {z}\right) }\right) }}_{z=\nu _{2}a}}\right| }^{2}}}}}}}}
\label{eq55}
\end{equation}

\noindent which confirms a rather complicated algebra.

The analysis shows that for the case of cylinder of small radius $ka\sqrt {%
\varepsilon - 1} < j_{0,1} \approx 2.4048$ only residue from term with $m=\pm 1$
will give contribution. As a result, the asymptote of decay rate of atom
placed at the surface of small dielectric fiber takes the following form $ka
\to 0$:

\begin{equation}
\begin{array}{l}
\left( {{\frac{{\gamma }^{total}}{{\gamma _{0}}}}}\right) _{\rho }=\left( {{%
\frac{{2\varepsilon }}{{\varepsilon +1}}}}\right) ^{2}+{\frac{{\varepsilon
^{2}\left( {\varepsilon -1}\right) \left( {15\varepsilon ^{2}+60\varepsilon
+2201-1680\left( {\gamma +\ln \left( {ka}\right) }\right) }\right) \left( {ka%
}\right) ^{2}}}{{300\left( {\varepsilon +1}\right) ^{3}}}}+O\left( {\left( {%
ka}\right) ^{4}}\right) \\ 
\\ 
+{\frac{{48\varepsilon ^{2}}}{{\left( {\varepsilon -1}\right) ^{2}\left( {ka}%
\right) ^{6}}}}\exp \left( {\ -{\frac{{2}}{{\left( {ka}\right) ^{2}}}}{\frac{%
{\varepsilon +1}}{{\varepsilon -1}}}+{\frac{{\varepsilon +1}}{{4}}}-2\gamma }%
\right)
\end{array}
\label{eq56}
\end{equation}

Here the first line describes the radiative losses while exponentially small
second line describes contribution of the main guided mode  $m=\pm 1$.

\section{Discussions  and Graphic illustrations}

We should note before all that asymptotes (\ref{eq49}) ,(\ref{eq53}) and (%
\ref{eq56}) do agree with quasistatic results (\ref{eq16}),(\ref{eq17})  in
the limit $ka\rightarrow 0$. This proves quasistatic calculations and confirms
complicated algebra of Sections IV,V.

Thus, to estimate radiative and nonradiative decay rates near dielectric
nanofiber with arbitrary complex permittivity 
$\varepsilon $  $\left(\rm{Re}\left( \varepsilon \right) >1\right)$
 one can use expressions (\ref{eq16},\ref{eq23}). 
To estimate  contribution of guided modes in nanofiber one
should use generalization of expressions obtained in Section V:

\[
\left( \frac{\gamma ^{guided}}{\gamma _{0}}\right) _{\rho }=%
{\rm Re}
\left( {\frac{{48\varepsilon ^{2}}}{{\left( {\varepsilon -1}\right)
^{2}\left( {ka}\right) ^{6}}}}\exp \left( {\ -{\frac{{2}}{{\left( {ka}%
\right) ^{2}}}}{\frac{{\varepsilon +1}}{{\varepsilon -1}}}+{\frac{{%
\varepsilon +1}}{{4}}}-2\gamma }\right) \right) 
\]

\[
\left( \frac{\gamma ^{guided}}{\gamma _{0}}\right) _{\varphi }=%
{\rm Re}
\left( {\frac{{48}}{{\left( {\varepsilon -1}\right) ^{2}\left( {ka}\right)
^{6}}}}\exp \left( {\ -{\frac{{2}}{{\left( {ka}\right) ^{2}}}}{\frac{{%
\varepsilon +1}}{{\varepsilon -1}}}+{\frac{{\varepsilon +1}}{{4}}}-2\gamma }%
\right) \right) 
\]

\begin{equation}
\left( \frac{\gamma ^{guided}}{\gamma _{0}}\right) _{z}=%
{\rm Re}
\left( {\frac{{12}}{{\left( {ka}\right) ^{4}}}}\exp \left( {\ -{\frac{{2}}{{%
\left( {ka}\right) ^{2}}}}{\frac{{\varepsilon +1}}{{\varepsilon -1}}}+{\frac{%
{\varepsilon +1}}{{4}}}-2\gamma }\right) \right)   \label{eq57}
\end{equation}

However, the asymptotics found do not allow us to determine the
region of their applicability. To find region of applicability of our
results we calculated the decay rates according to full formulae and compare
them with asymptotes obtained in previous Section. The results of comparison
are presented in Figs. 4-12.

First of all from Figs. 4,7,10 one can see that our asymptotic expansions
are good enough for $ka<0.4(\varepsilon =3)$ . When the dielectric
permittivity is increased the region of applicability of longwave asymptotes
is reduced. For $\varepsilon =10$ (Figs. 5,8,11) applicability region is $%
ka<0.1$, while for $\varepsilon =30$ (Figs. 6,9,12) applicability region is 
$ka<0.05$. One can suppose that generally our asymptotics are good  for

\begin{equation}
ka<1/\varepsilon   \label{eq58}
\end{equation}

\noindent The restriction  is  more rigid than it may appear  from cursory examination
($\sqrt{\varepsilon }ka<1)$.

This region corresponds to rather thin nanofiber. For example, for $%
\varepsilon =3$ fiber radius should be about $\lambda /20$! The case of
large dielectric permittivity should be treated carefully because the limits 
$ka\rightarrow 0$ and $\varepsilon \rightarrow \infty $ do not commute. To
investigate the case of ideal conducting nanowire $(|\varepsilon
|\rightarrow \infty )$, one should take limit of general expressions (\ref
{eq48}), (\ref{eq52}), (\ref{eq54}) and only then investigate asymptotics
for $ka\rightarrow 0$. The analogous situation takes place in the case of
planar interface \cite{Eberlein} or for atom near prolate nanospheroid \cite
{Klimov2,Klimov3}.

In the region $1/\varepsilon<ka <1/\sqrt{\varepsilon}$ our asymptotics give satisfactory approximation. The rest of nanofiber radii,

\begin{equation}
1/\sqrt{\varepsilon} <ka<2.4/\sqrt{\varepsilon -1}  \label{eq59}
\end{equation}

\noindent should be analyzed numerically (see Figs.4-12). From these figures
one can see that for large enough nanofibers (\ref{eq59}), the decay rates
increase with increasing of dielectric permittivity. The most substantial
enhancement is observed for $\rho $ and z orientations, where enhancement of
total decay rates can reach value about 35, and 23 for $\varepsilon =30$
(Figs.6,9). The most important feature of $\rho $- and $z$- orientations of
dipole momentum is very efficient excitation of guided modes. On the
contrary, the influence of guided modes on spontaneous emission of 
$\varphi $-oriented dipole is rather small.

To trace the influence of guided mode on total decay rate we plot the ratio
of decay rate into guided modes to total decay rate

\begin{equation}
\beta ={\frac{{\left( {{\frac{{\gamma }}{{\gamma _{0}}}}}\right) ^{guided}}}{%
{\left( {{\frac{{\gamma }}{{\gamma _{0}}}}}\right) ^{guided}+\left( {{\frac{{%
\gamma }}{{\gamma _{0}}}}}\right) ^{radiative}}}}  \label{eq60}
\end{equation}
In (\ref{eq60}) we neglect  a contribution from the nonradiative processes.

 This quantity is very important for determining the laser threshold \cite
{Yamamoto}. From Figs. 13-15 it is seen that the spontaneous emission
coupling efficiency $\beta $ is rather high for $\rho $ - and z - orientations
even in the case of monomode nanofibers! It is interesting that the
asymptote describes the coupling efficiency well in rather wide region
(Fig.13). The dipoles with $\varphi$ - orientations of momentum have a
small spontaneous emission coupling efficiency (Fig.14)

Finally, in  Fig.16 the total decay rates near a nanofiber are compared
with decay rates near an ideally conducting cylinder. From the figure one can
see that for large enough dielectric permittivity the decay rate near
nanofiber can be greater than decay rate near ideally conducting cylinder.
Again, that effect is due to excitation of principal mode in a nanofiber.

\section{Conclusion}

In the present paper the decay rates of an excited atom placed near a
dielectric fiber are considered. The main attention was paid to the case of
cylinder with radius which is small in comparison with radiation wavelength
(nanofiber), $ka<2.4/\sqrt{\varepsilon -1}$. The decay rates are found
within quasistatic as well as full electrodynamic approaches. It is proved
that quasistatic approximation works well for a nanofiber with $ka<1/\varepsilon$. In
contrast to quasistatic solution the exact one has additional terms from
guided modes, which exist even for nanofiber of arbitrarily small radius.
However the contributions from such modes decreases exponentially when
cylinder radius tends to zero. For large enough nanofiber, ${\rm 1}{\rm /}%
\varepsilon <ka<2.4/\sqrt{\varepsilon -1}$, the influence of guided modes on
the decay rate is substantial.

The results obtained can be useful as for estimation of decay rates and for
understanding of interplay between different decay channels. The results
obtained are in agreement with those for an atom placed near dielectric or
metallic nanospheroid \cite{Klimov2,Klimov3}.

In the present paper we pay attention to the case of dielectric nanofiber with
positive dielectric permittivity. However, our results can be applied to
investigation of decay rates near metallic nanowire with negative dielectric
constant. In the case of nanowires the quasistatic expressions (\ref{eq16})
remain valid for description of radiative losses, but one should add to them
a contribution, which is due to excitation of symmetric guided modes. The
detailed analysis of decay rates of an atom placed near nanowires will be
presented elsewhere \cite{Klimov4}.

\begin{center}
{\bf Acknowledgements}
\end{center}

The authors thank the Russian Basic Research Foundation (V.K.) , Center
``Integration'' and Centre National De La Recherche Scientifique for
financial support of this work.

\newpage 
\begin{figure}
\epsfxsize 5 in
\centerline{\epsfbox{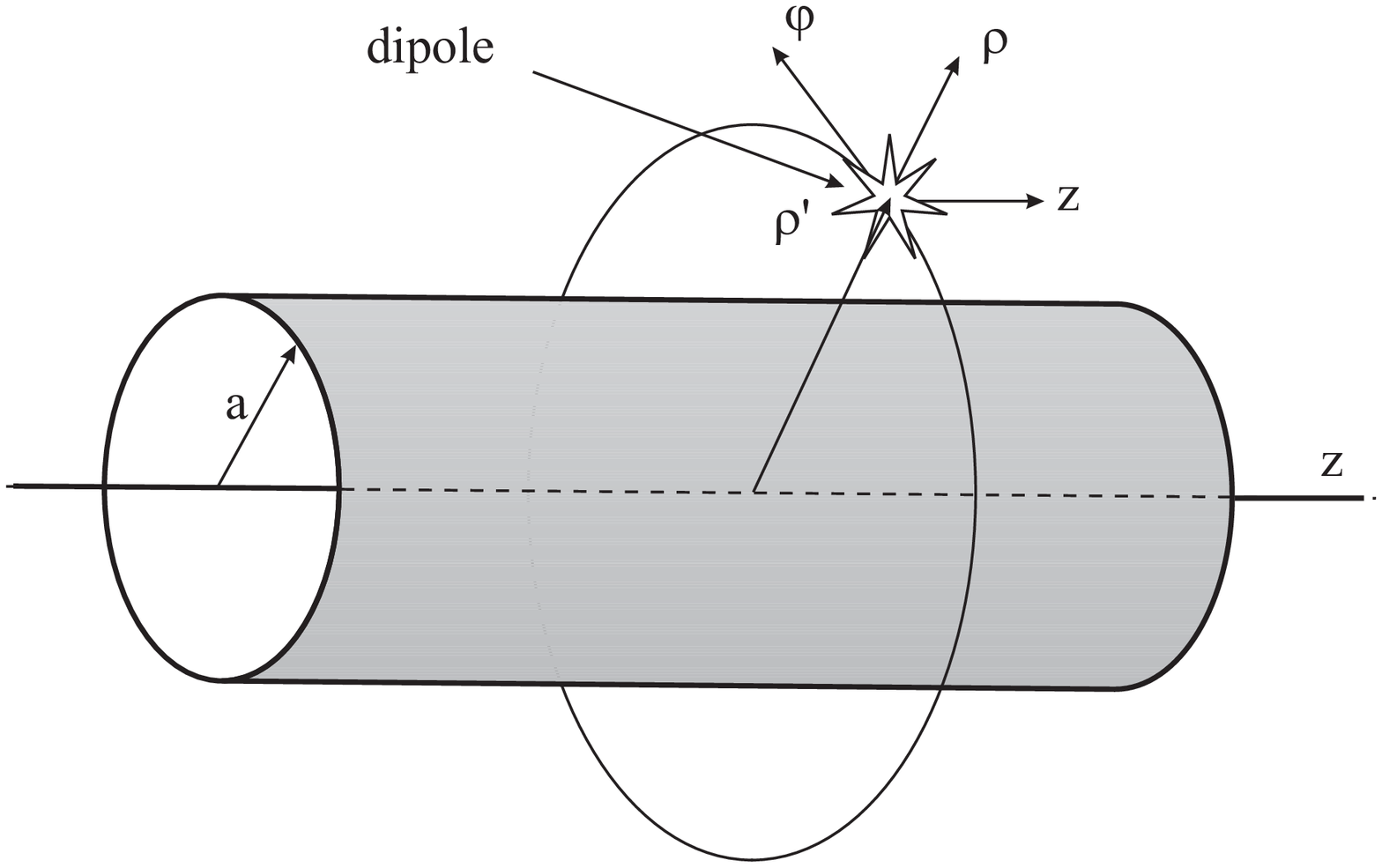}}
\caption{Geometry of the problem.}
\end{figure}
\newpage
\begin{figure}
\epsfxsize 4 in
\centerline{\epsfbox{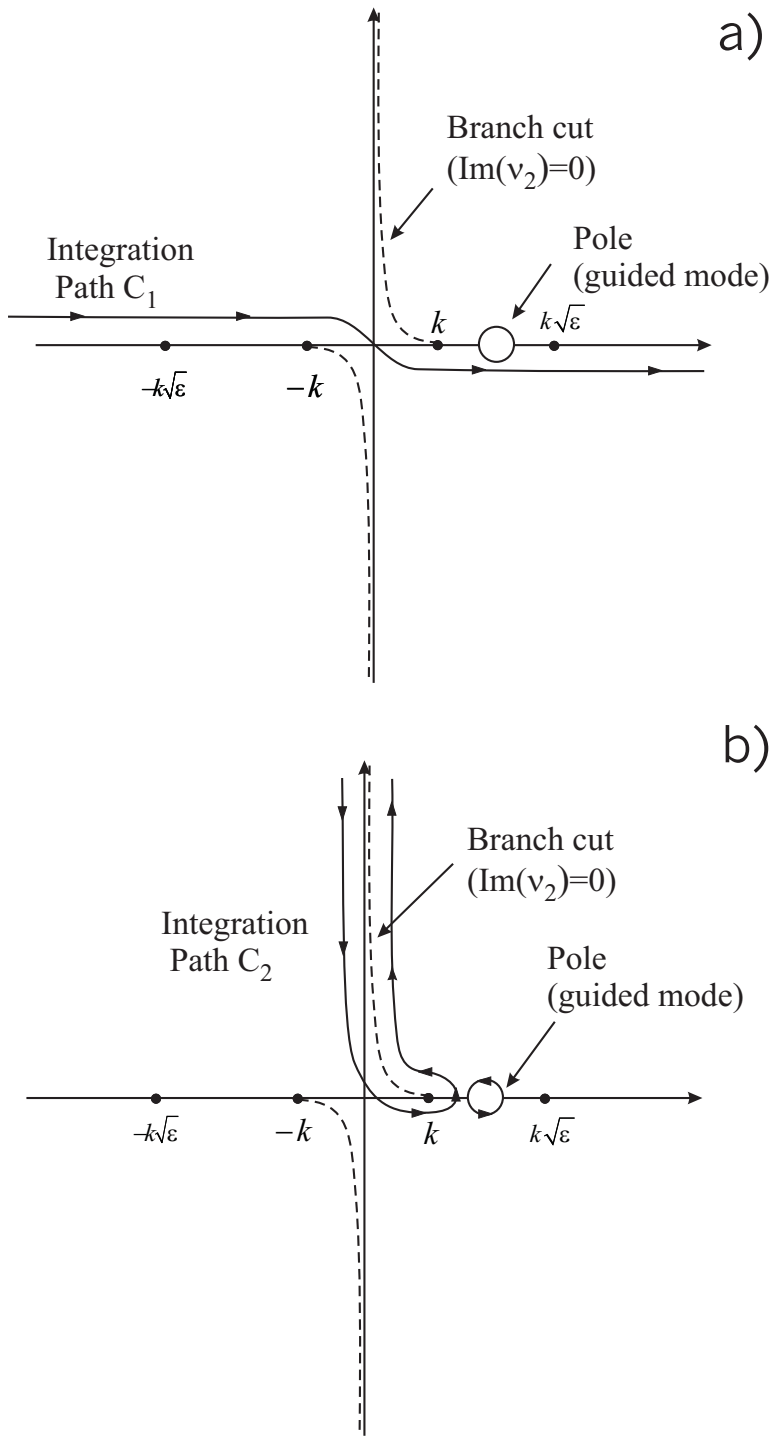}}

\caption{Integration path for contour integral over longitudinal wavenumber $h$.}
\end{figure}

\newpage

\begin{figure}
\epsfxsize 4 in
\centerline{\epsfbox{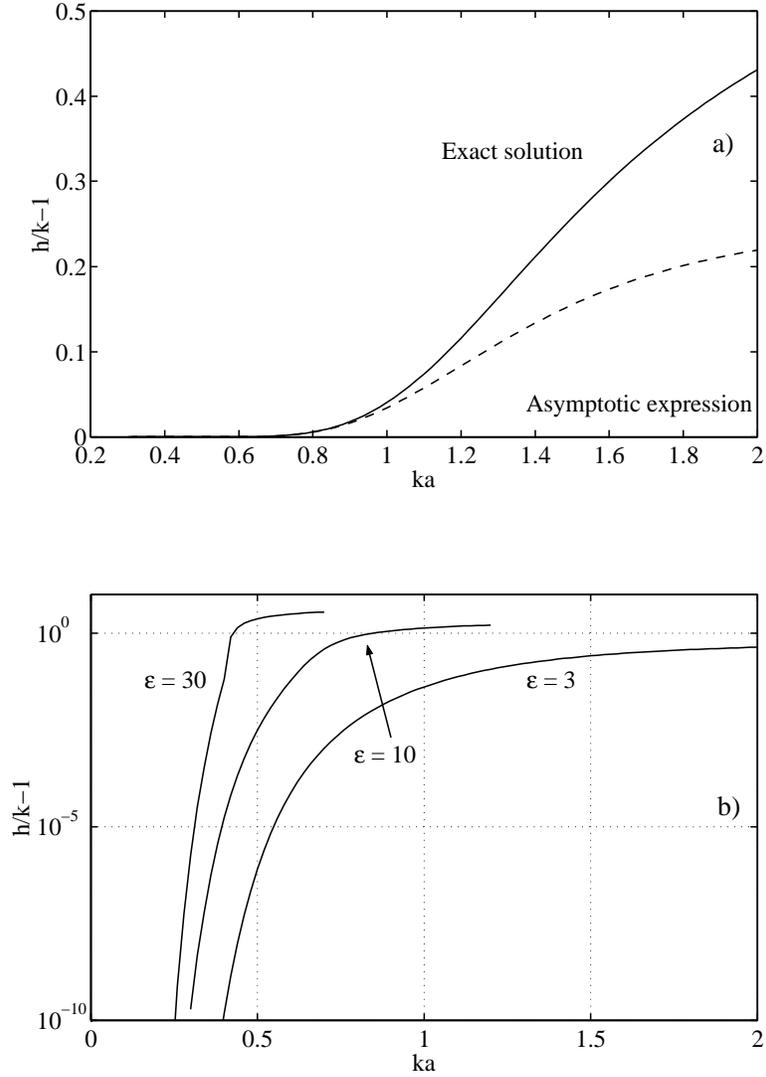}}
\caption{Dependence of longitudinal wavenuber of main mode $(h/k)_{01}$ on radius of
fiber. a)exact and asymptotic solutions for $\epsilon=3$, b)exact solutions for  %
$\epsilon=3,10,30$.} 
\end{figure}

\newpage

\begin{figure}
\epsfxsize 4 in
\centerline{\epsfbox{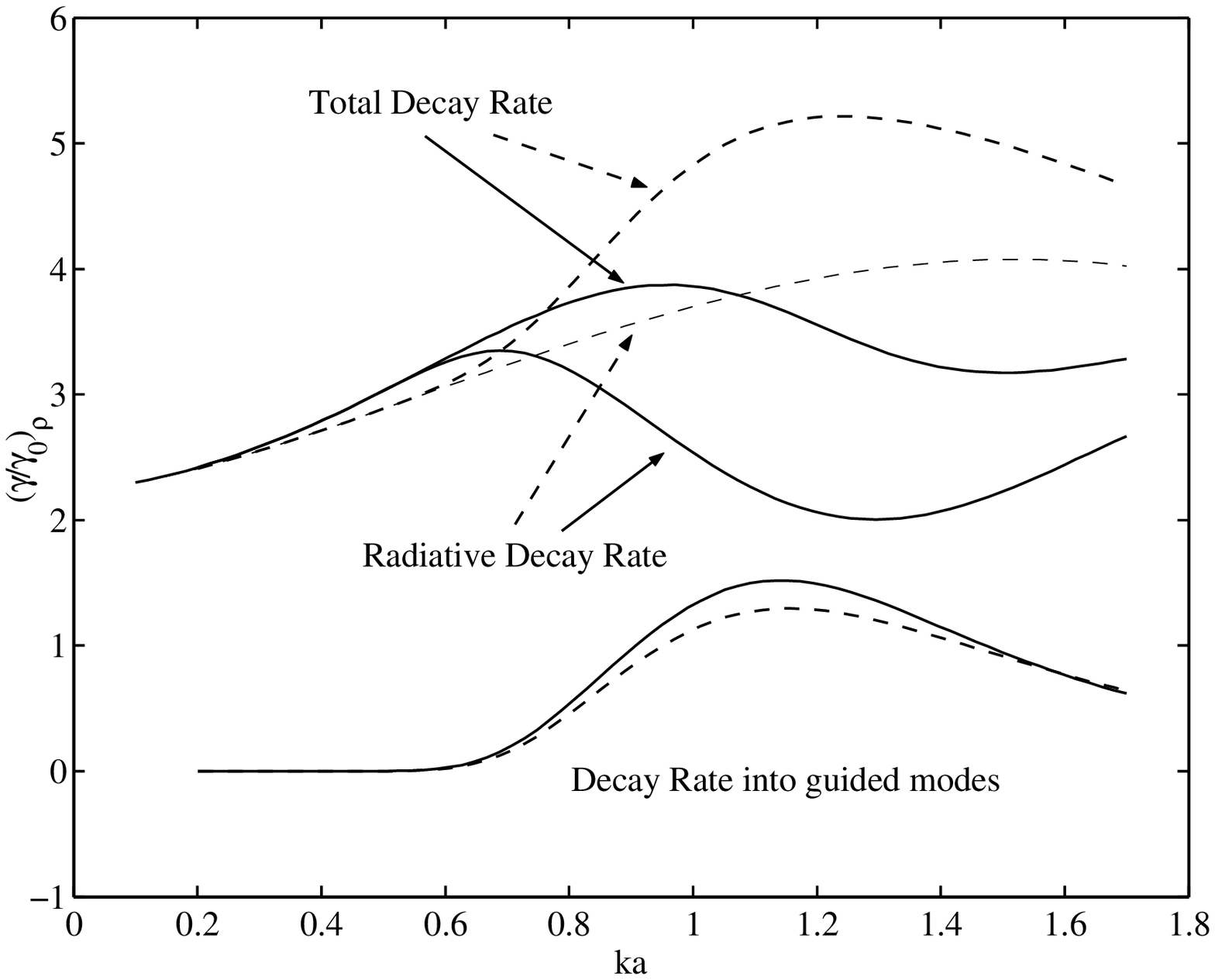}}
\caption{Relative decay rates of an atom with $\rho $ orientation as a
function of radius of fiber $ka$ [Eq.(\ref{eq54}), solid lines]) and its %
 asymptotic expression [Eq.(\ref{eq56}), dashed lines]. 
(atom on the surface, $\varepsilon = 3$)}
\end{figure}

\begin{figure}
\epsfxsize 4 in
\centerline{\epsfbox{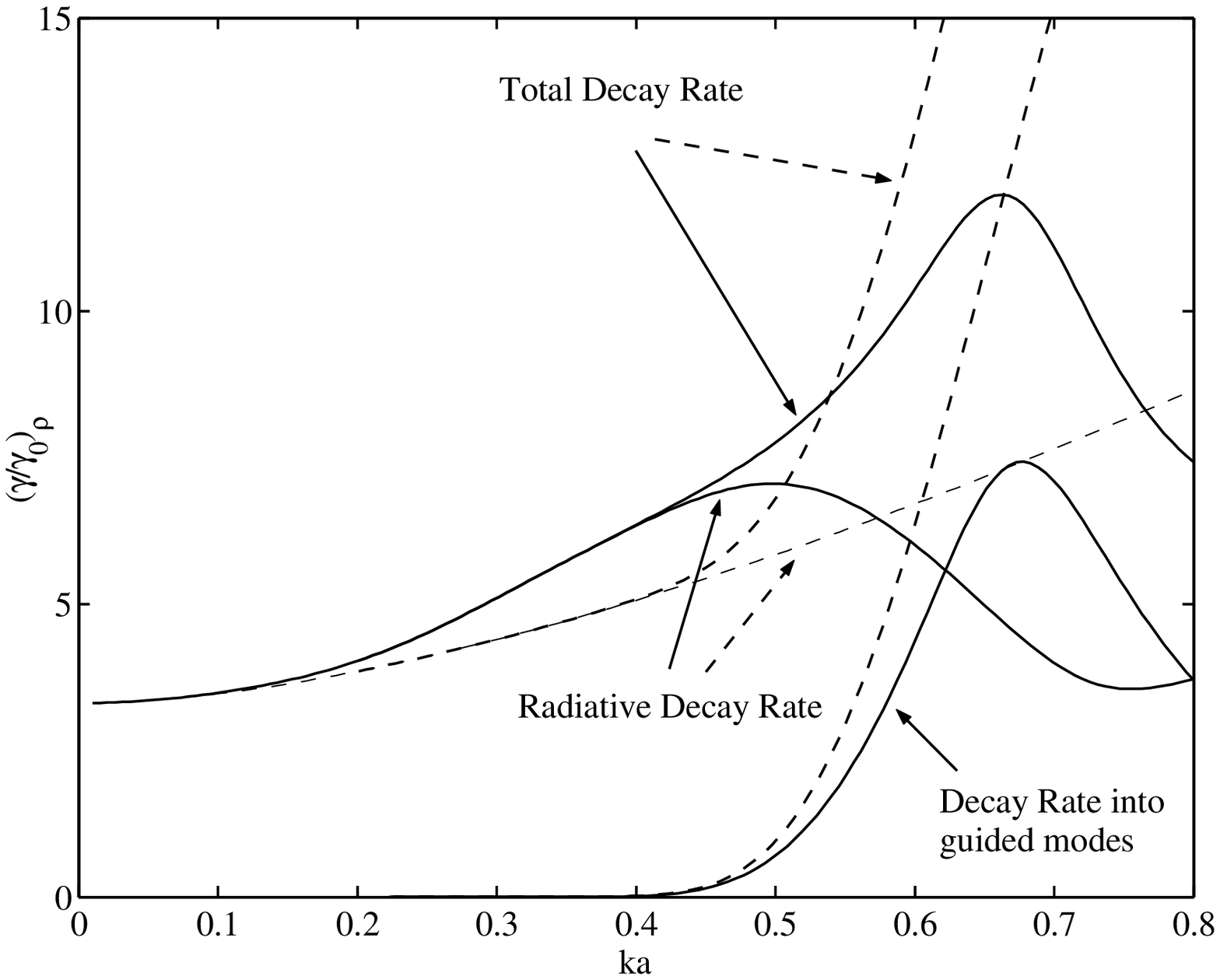}}
\caption{Relative decay rates of an atom with $\rho $ orientation as a
function of radius of fiber $ka$ [Eq.(\ref{eq54}), solid lines]) and its %
 asymptotic expression [Eq.(\ref{eq56}), dashed lines]. 
(atom on the surface, $\varepsilon = 10$)}
\end{figure}
\newpage

 \begin{figure}
\epsfxsize 4 in
\centerline{\epsfbox{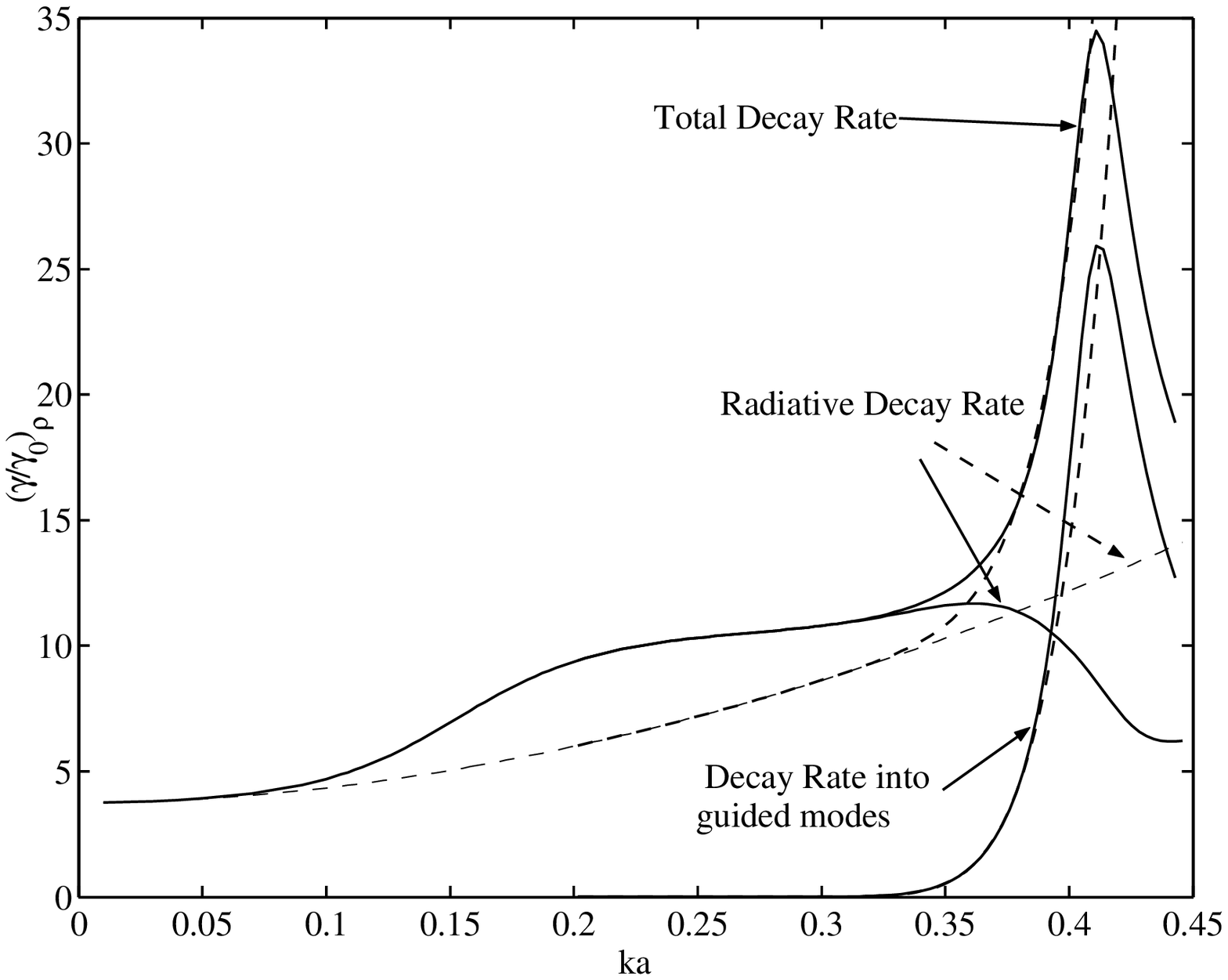}}
\caption{Relative decay rates of an atom with $\rho $ orientation as a
function of radius of fiber $ka$ [Eq.(\ref{eq54}), solid lines]) and its %
 asymptotic expression [Eq.(\ref{eq56}), dashed lines]. 
(atom on the surface, $\varepsilon = 30$)}
\end{figure}

\begin{figure}
\epsfxsize 4 in
\centerline{\epsfbox{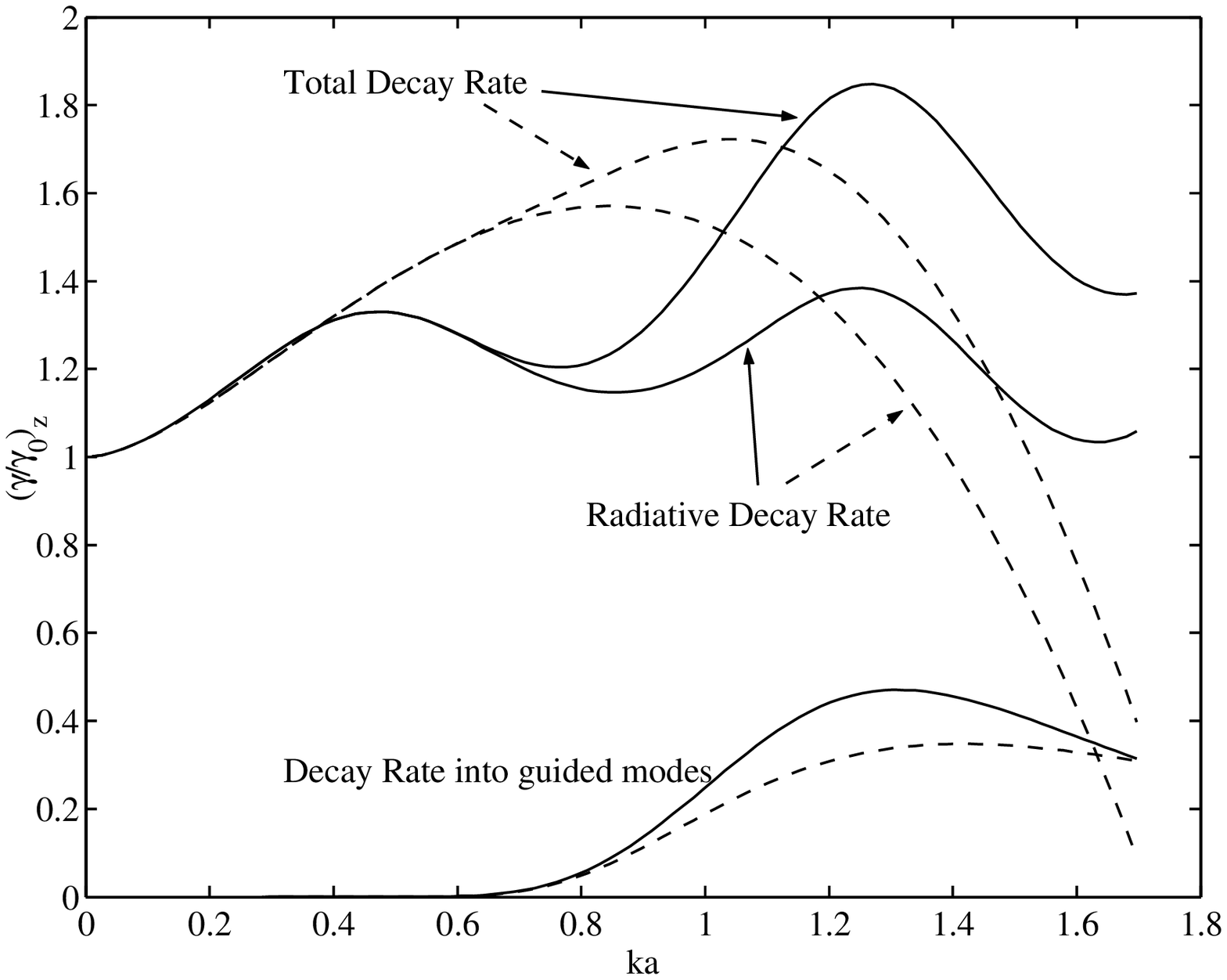}}
\caption{Relative decay rates of an atom with $ z$ orientation as a
function of radius of fiber $ka$ [Eq.(\ref{eq48}), solid lines]) and its %
 asymptotic expression [Eq.(\ref{eq49}), dashed lines]. 
(atom on the surface, $\varepsilon = 3$)}
\end{figure}
\newpage

 \begin{figure}
\epsfxsize 4 in
\centerline{\epsfbox{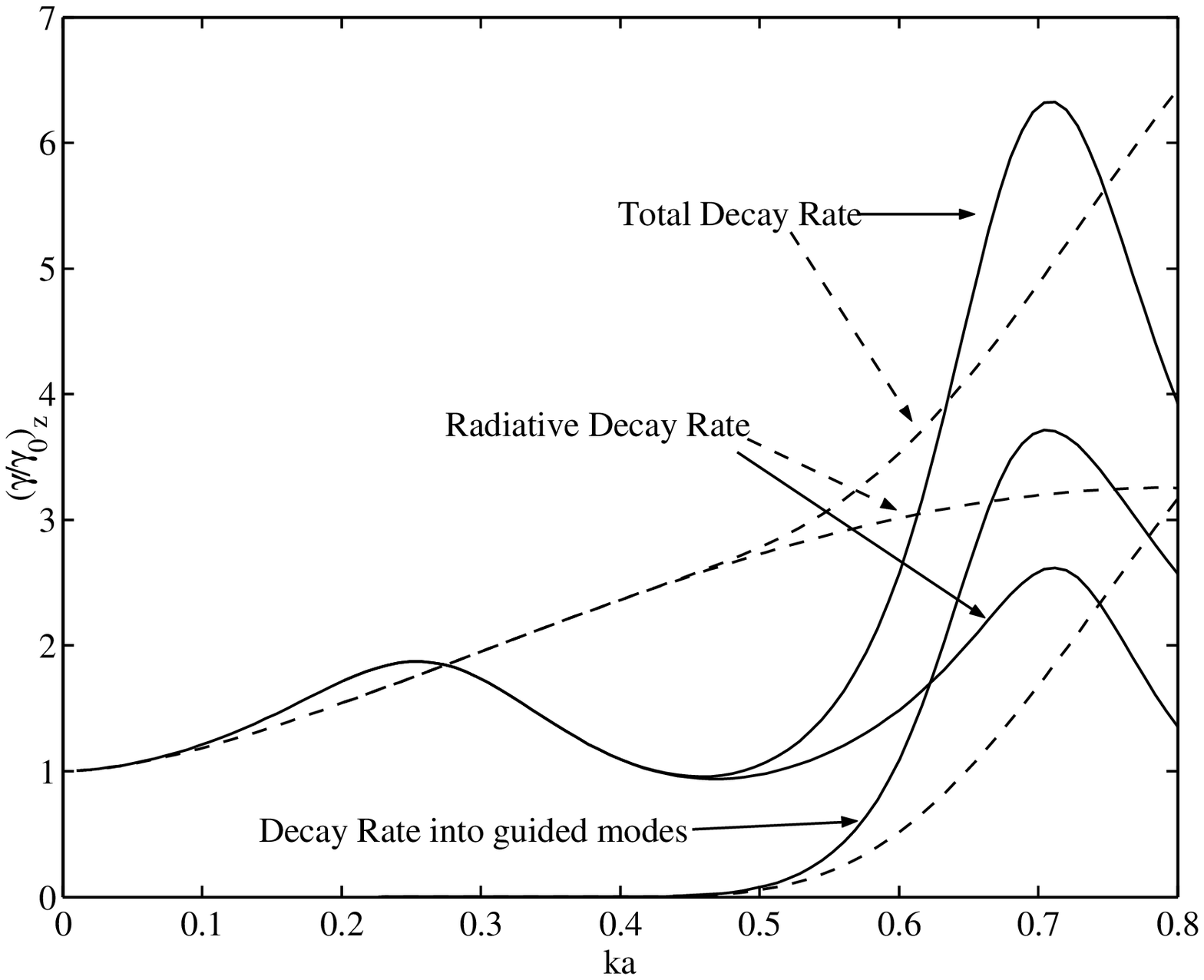}}
\caption{Relative decay rates of an atom with $ z$ orientation as a
function of radius of fiber $ka$ [Eq.(\ref{eq48}), solid lines]) and its %
 asymptotic expression [Eq.(\ref{eq49}), dashed lines]. 
(atom on the surface, $\varepsilon = 10$)}
\end{figure}

\begin{figure}
\epsfxsize 4 in
\centerline{\epsfbox{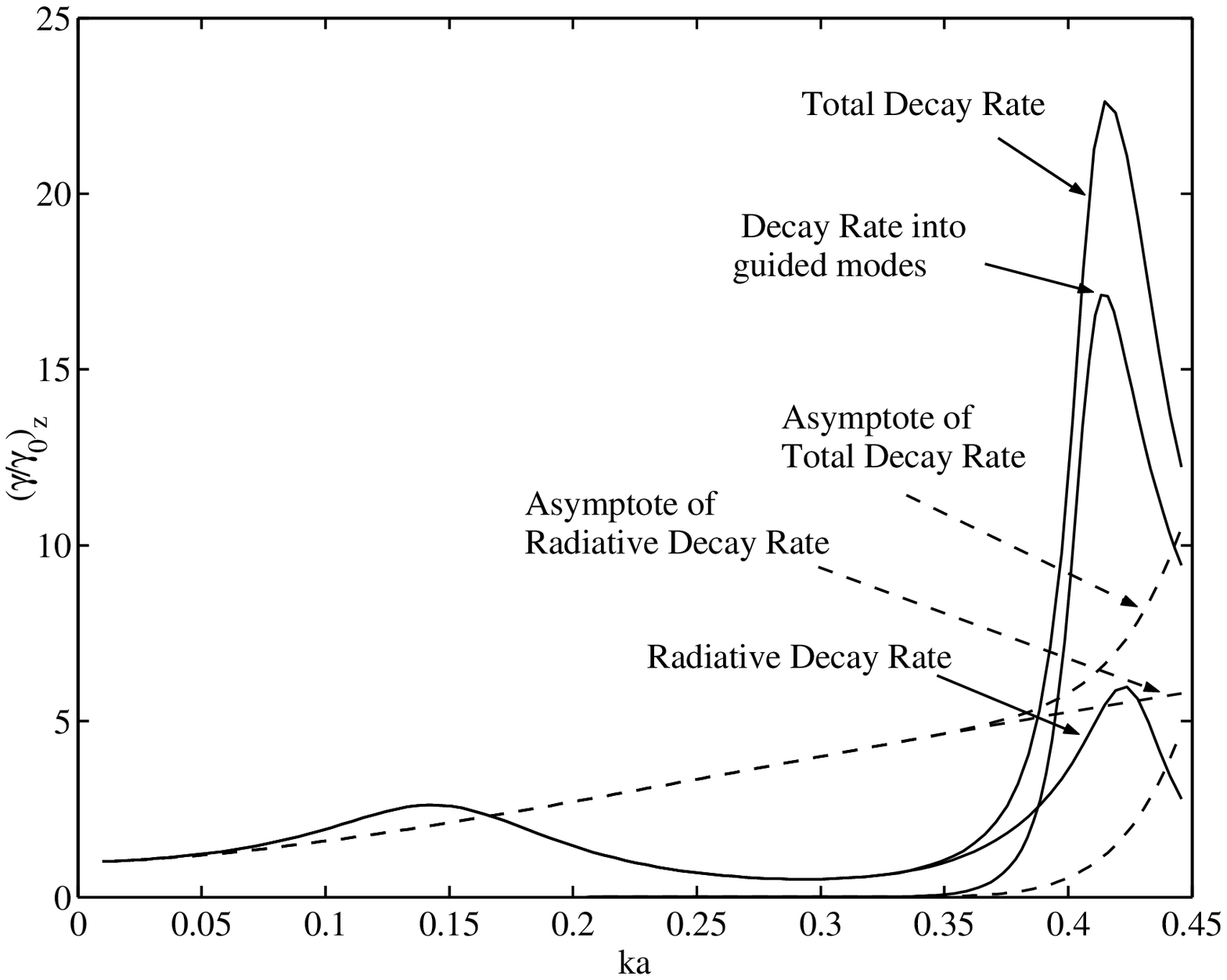}}
\caption{Relative decay rates of an atom with $ z$ orientation as a
function of radius of fiber $ka$ [Eq.(\ref{eq48}), solid lines]) and its %
 asymptotic expression [Eq.(\ref{eq49}), dashed lines]. 
(atom on the surface, $\varepsilon = 30$)}
\end{figure}
\newpage

 \begin{figure}
\epsfxsize 4 in
\centerline{\epsfbox{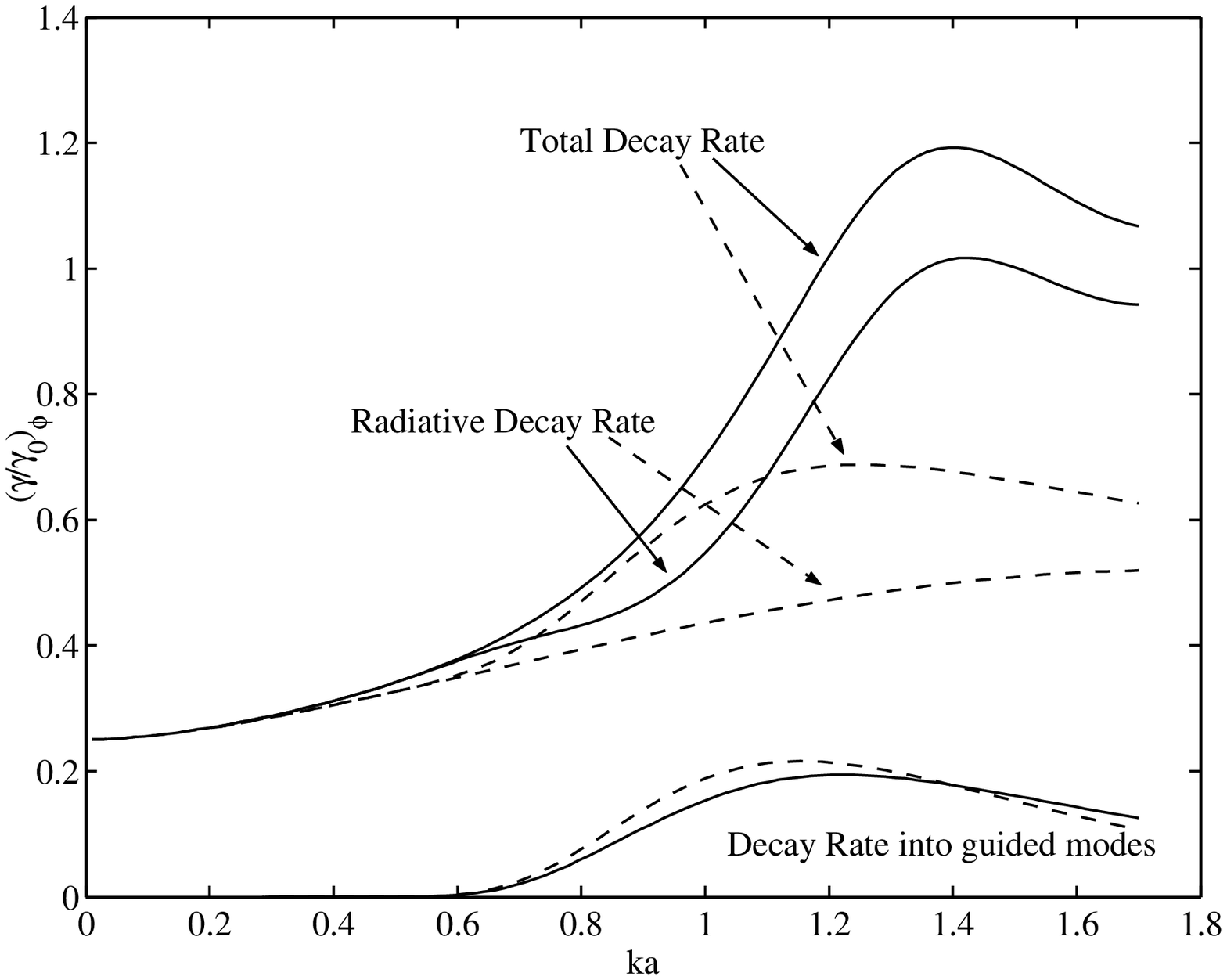}}
\caption{Relative decay rates of an atom with $\varphi $ orientation as a
function of radius of fiber $ka$ [Eq.(\ref{eq52}), solid lines]) and its %
 asymptotic expression [Eq.(\ref{eq53}), dashed lines]. 
(atom on the surface, $\varepsilon = 3$)}
\end{figure}

\begin{figure}
\epsfxsize 4 in
\centerline{\epsfbox{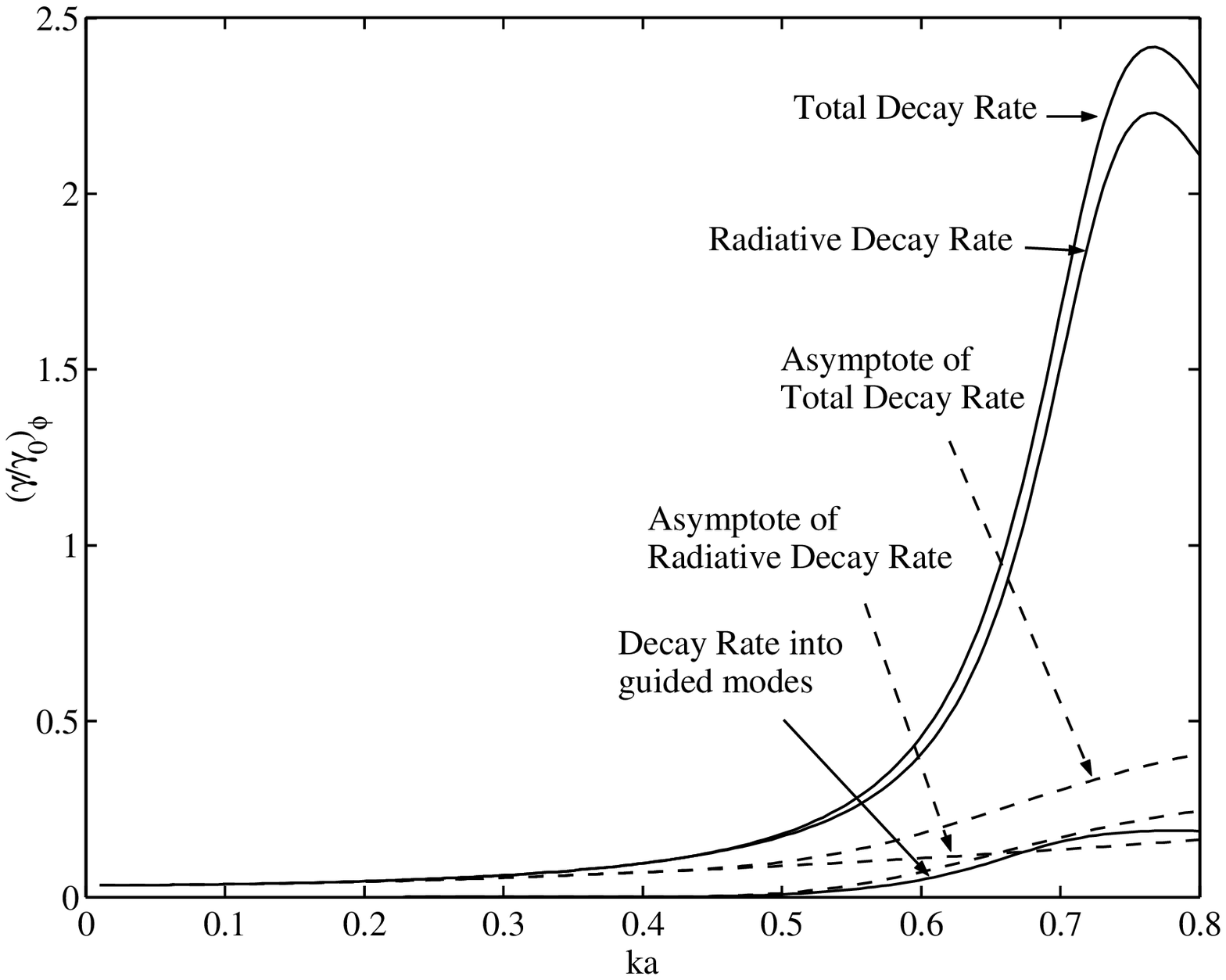}}
\caption{Relative decay rates of an atom with $\varphi $ orientation as a
function of radius of fiber $ka$ [Eq.(\ref{eq52}), solid lines]) and its %
 asymptotic expression [Eq.(\ref{eq53}), dashed lines]. 
(atom on the surface, $\varepsilon = 10$) }
\end{figure}
\newpage

 \begin{figure}
\epsfxsize 4 in
\centerline{\epsfbox{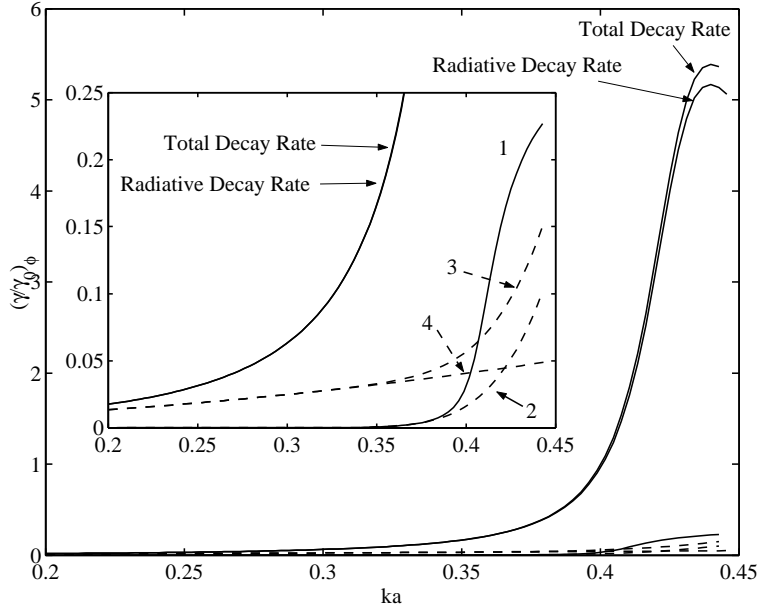}}
\caption{  Relative decay rates of an atom with $\varphi $ orientation as a
function of radius of fiber $ka$ [Eq.(\ref{eq52}), solid lines]) and its %
 asymptotic expression [Eq.(\ref{eq53}), dashed lines]. 
(atom on the surface, $\varepsilon = 30$). On the inset: 1
- decay rate into guided modes, 2 - asymptotic expression of decay rate into guided
modes, 3 - asymptotic expression of total decay rate, 4 - asymptotic expression of radiative decay
rate}
\end{figure}

\begin{figure}
\epsfxsize 4 in
\centerline{\epsfbox{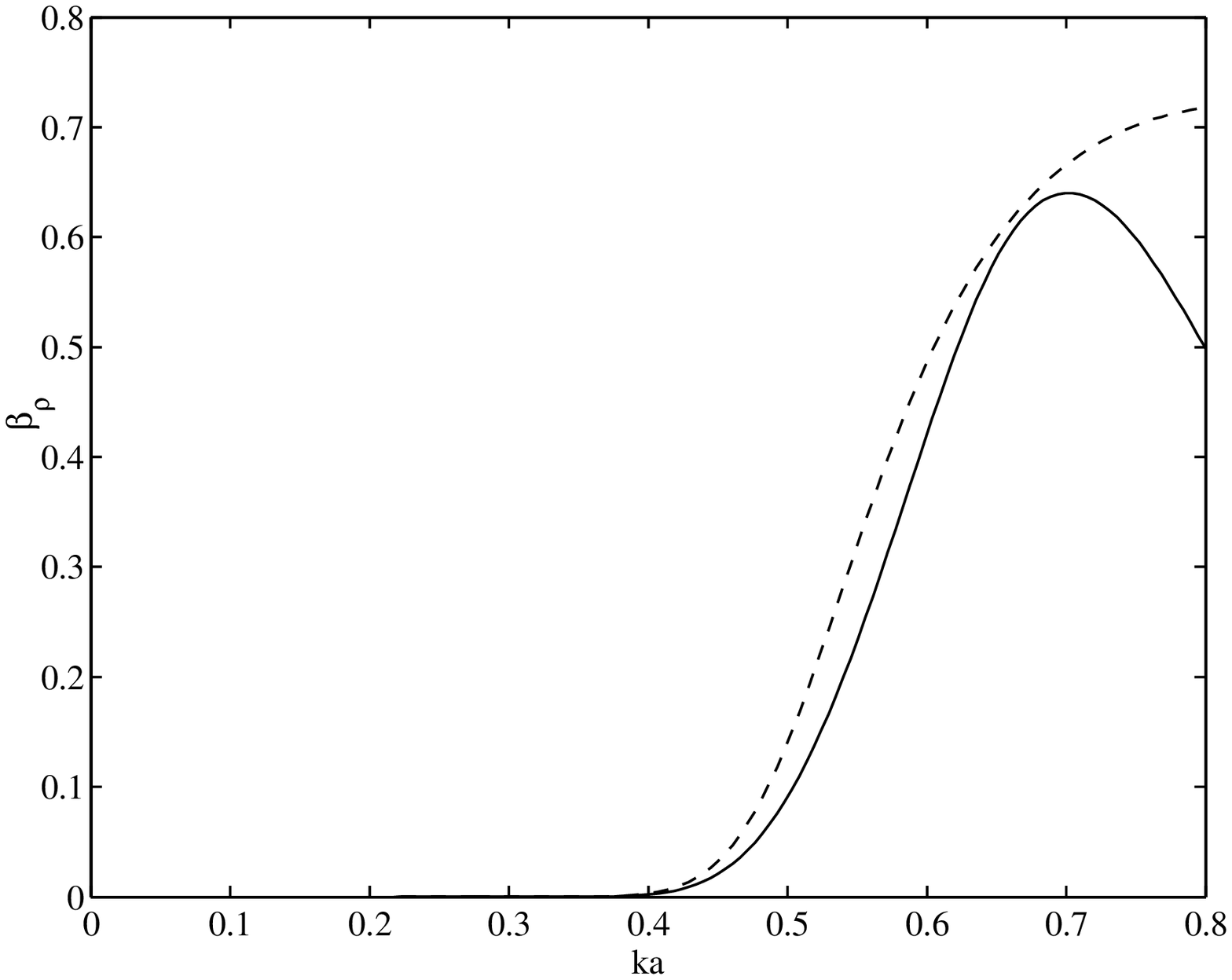}}
\caption{ The spontaneous emission coupling efficiency [Eq.(\ref{eq60})] as a
function of radius of fiber ka (solid lines ) and it asymptotic expressions (dashed
lines) ($\rho $ - orientation, $\varepsilon = 10$ )}
\end{figure}
\newpage

 \begin{figure}
\epsfxsize 4 in
\centerline{\epsfbox{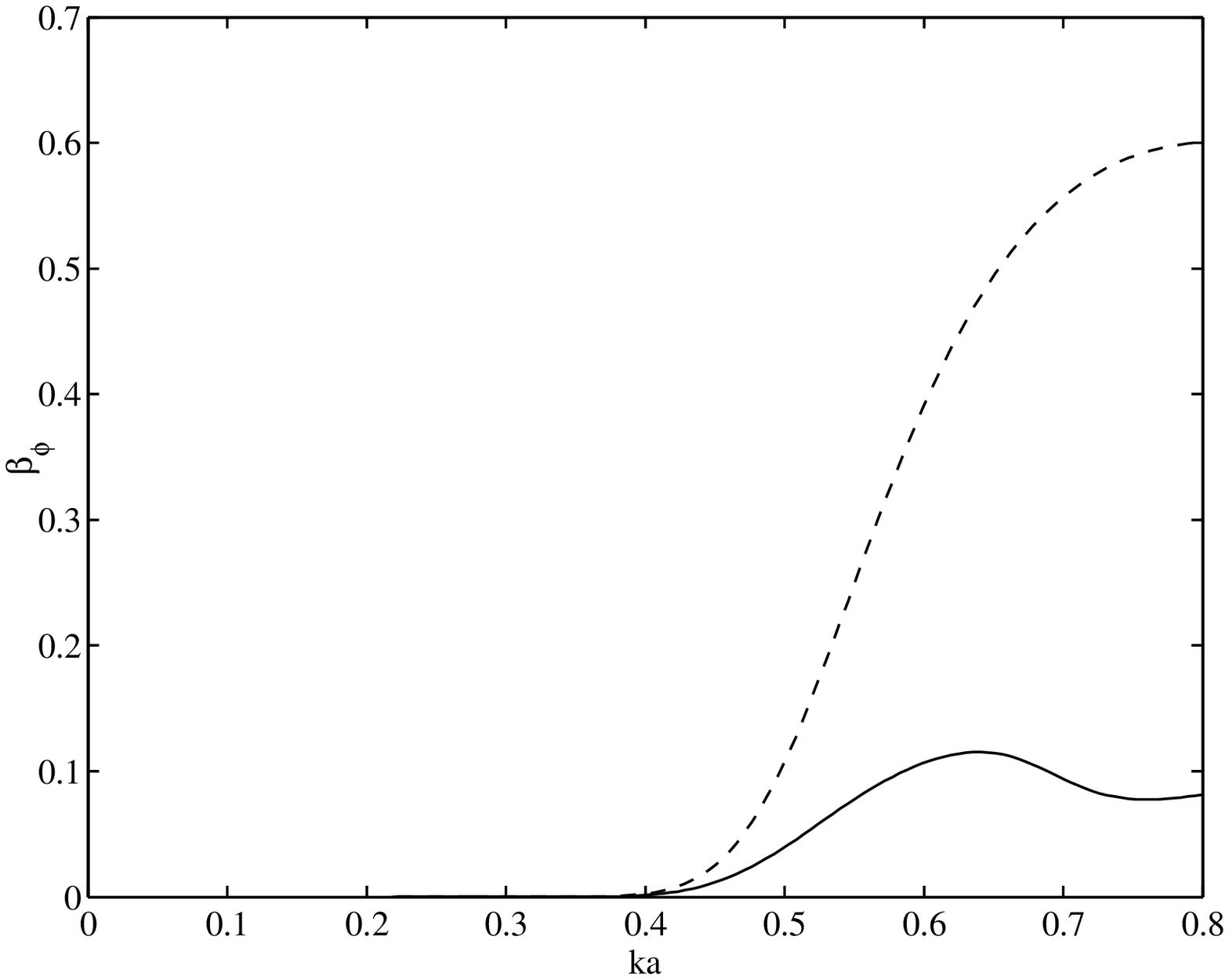}}
\caption{  The spontaneous emission coupling efficiency [Eq.(\ref{eq60})] as a
function of radius of fiber ka (solid lines ) and it asymptotic expressions (dashed
lines) ( $\varphi $ - orientation, $\varepsilon = 10$ ).}
\end{figure}

\begin{figure}
\epsfxsize 4 in
\centerline{\epsfbox{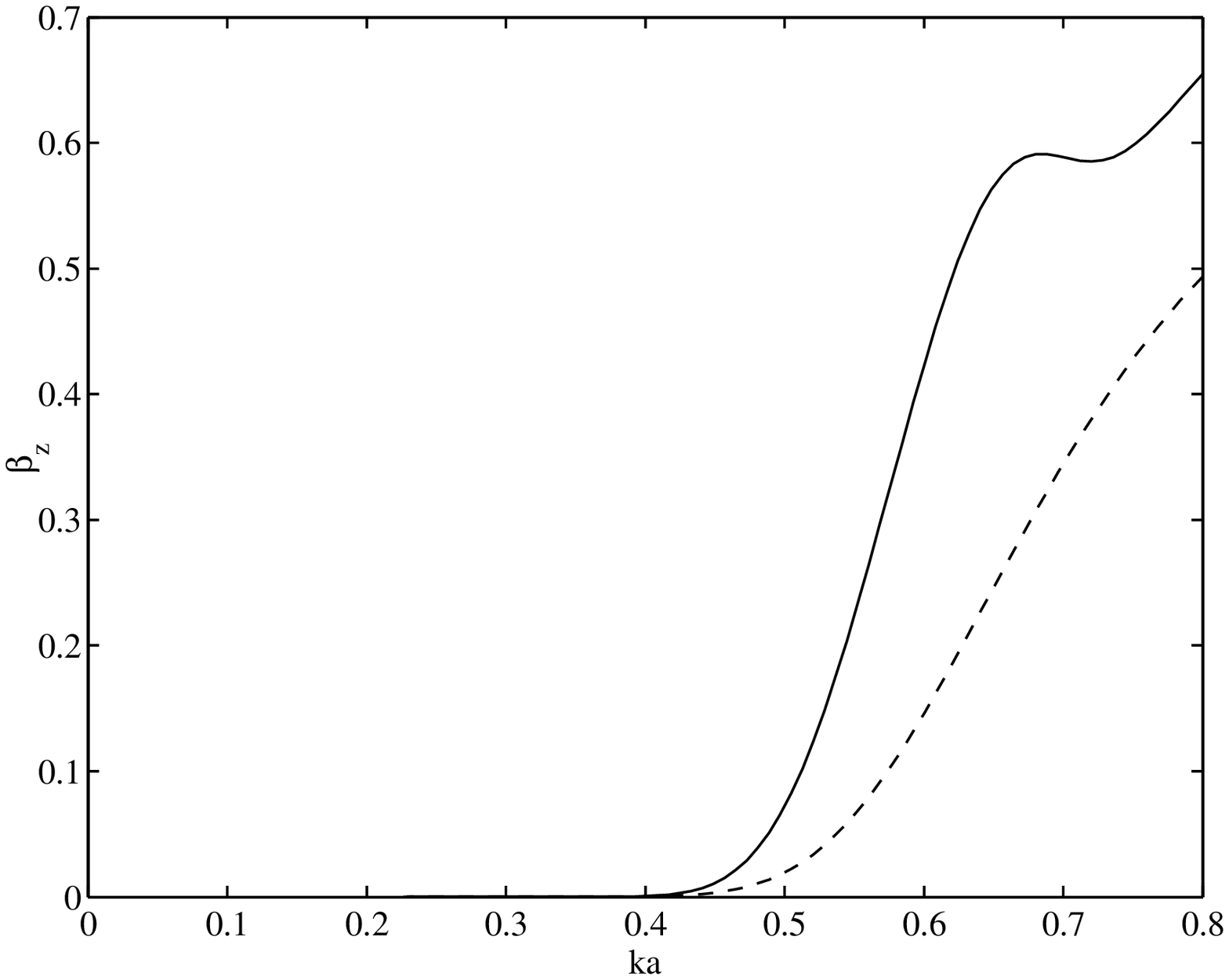}}
\caption{ The spontaneous emission coupling efficiency [Eq.(\ref{eq60})] as a
function of radius of fiber ka (solid lines ) and it asymptotic expressions (dashed
lines) (z - orientation, $\varepsilon = 10$ ).}
\end{figure}
\newpage

\begin{figure}
\epsfxsize 4 in
\centerline{\epsfbox{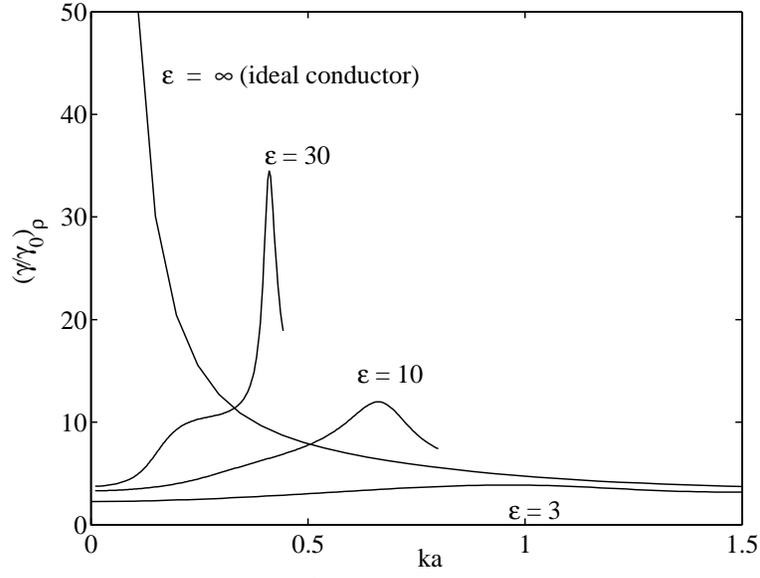}}

\caption{  Relative linewidth [Eq.(\ref{eq54})] of dipole transitions with $\rho $
orientation as a function of fiber radius ka for different $\varepsilon =
3, 10, 30 $ and for ideally conducting cylinder (atom on the surface).}
\end{figure}

\end{document}